\documentclass[submission, Phys]{SciPost}

\hypersetup{
    colorlinks,
    linkcolor={red!50!black},
    citecolor={blue!50!black},
    urlcolor={blue!80!black}
}

\usepackage[bitstream-charter]{mathdesign}
\DeclareSymbolFont{usualmathcal}{OMS}{cmsy}{m}{n}
\DeclareSymbolFontAlphabet{\mathcal}{usualmathcal}

\begin{document}

\begin{center}{\Large \textbf{
Non-equilibrium Sachdev-Ye-Kitaev model with quadratic perturbation\\
}}\end{center}

\begin{center}
Aleksey V. Lunkin\textsuperscript{1,2,3$\star$} and
Mikhail V. Feigel'man\textsuperscript{2,4} 
\end{center}

\begin{center}
{\bf 1} Skolkovo Institute of Science and Technology, 143026 Skolkovo, Russia
\\
{\bf 2} L. D. Landau Institute for Theoretical Physics, Kosygin Str.~2, Moscow 119334, Russia
\\
{\bf 3} HSE University, Moscow, Russia
\\
{\bf 4} Moscow Institute of Physics and Technology, Moscow 141700, Russia
\\
${}^\star$ {\small \sf alunkin@itp.ac.ru}
\end{center}

\begin{center}
\today
\end{center}


\section*{Abstract}
{\bf
We consider a non-equilibrium generalization of the  mixed SYK$_4$+SYK$_2$ model
and calculate the energy dissipation rate
$W(\omega)$ that results  due to periodic modulation of random quadratic matrix elements  with  a frequency $\omega$. 
We find that $W(\omega)$ possesses a peak at $\omega$ close to the polaron energy spliting $\omega_R$
found  recently in~\cite{lunkin2020perturbed}, 
demonstrating  physical significance of this energy scale. 
Next, we study the effect of energy pumping with a finite amplitude  at the resonance frequency $\omega_R$
and calculate, in presence of this pumping, non-equilibrium dissipation rate due to low-frequency parameteric modulation. 
We found unusual phenomenon similar to "dry friction" in presence of pumping.
}

\vspace{10pt}
\noindent\rule{\textwidth}{1pt}
\tableofcontents\thispagestyle{fancy}
\noindent\rule{\textwidth}{1pt}
\vspace{10pt}

\section{Introduction}
\label{sec:intro}
The Sachdev-Ye-Kitaev model~\cite{Kitaev2015,kitaev2018soft,maldacena2016remarks,bagrets2016sachdev} modified by random
quadratic terms in the Hamiltonian presents a valuable starting point to develop a theory of strongy correlated
electron systems.  Recent results~\cite{lunkin2020perturbed} demonstrate an interesting interplay between
soft-mode fluctuations (which dominates the infra-red behavior of the pure SYK$_4$ model) and  SYK$_2$ terms
of  moderate magnitude.  Namely, it was found that in  presence of SYK$_2$ terms the soft-mode fluctuations become
suppressed in a wide temperature range, and can be described by a kind of nearly Gaussian action. 
The crucial role in this picture is played by the bound-polaron solution obtained in~\cite{lunkin2020perturbed} 
for some collective Bose field. 
The Liouville quantum mechanics approach~\cite{bagrets2016sachdev,kitaev2018soft} to the pure SYK$_4$ model
can be recast into the form of the functional integral over the same Bose field, but in this case bound-states
are absent and fluctuations are strong and non-Gaussian.
We have shown~~\cite{lunkin2020perturbed}  that quadratic  perturbation lead to the formation of the polaron  bound-state
and thus to suppression of fluctuations at low energies.

In the present Letter we extend the approach of Ref.~\cite{lunkin2020perturbed} into a broad field of non-equilibrium 
problems described by Keldysh functional integral methods.
We  study the simplest physical quantity that may shed some light on the physical significance of
the polaron bound-state.
Namely, we calculate the power dissipated in the system due to time-dependent modulation of the quadratic part of 
the Hamiltonian.  In other words, we propose here generalization of the approach 
well-known~\cite{gorkov1965zh,efetov1983supersymmetry,simons1993universal,skvortsov2004energy} for non-interacting random Fermi-systems, where  dissipation due to time-dependent periodic  perturbation can be expressed in terms of parametric statistics of a Wigner-Dyson
random-matrix ensemble. 

\section{The model}
The Hamiltonian of the model has the form:
\begin{equation}
H=\frac{1}{4!}\sum_{ijkl} J_{ijkl}\chi_i\chi_j\chi_k\chi_l+\frac{i}{2}\sum_{ij} \Gamma_{ij} (1+\Phi(t))\chi_i \chi_j.
\end{equation}
Here $J_{ijkl}$ and $\Gamma_{ij}$ are random Gaussian antisymmetric tensors. Their average values vanish, 
while $ \overline{J_{ijkl}^2} = \frac{3!J^2}{N^3} $
and  $ \overline{\Gamma_{ij}^2}= \frac{\Gamma^2}{N}$;
here $\chi_l$  is the Majorana fermion operator, index $l\in (1,N)$ and $N$ is the full number of sites available for fermions. 
Our aim is to calculate the energy dissipation rate $W(\omega)$  caused by the
periodic modulation described by $\Phi(t)$. Below we will show that pumping at a certain frequency $\omega_P$ (its value
will be specified below)
can  change the properties of the model crucially, so it is useful to consider
$\Phi(t)=A\cos(\omega_P t)+f(t)$ and to study the energy dissipation rate due to the application of the field 
$f(t) \propto e^{-i\omega t}$.  
To begin with, we make several technical notes about our calculation and introduce few useful notations.

We  use dimensionless time $x\equiv\frac{2\pi t}{\beta}$  and dimensional frequency $\Omega=\frac{\omega \beta}{2\pi}$. As we plan to study non-equilibrium properties of the  model with strong interaction, we need to use Keldysh formalism~\cite{kamenev2011field}. The Green function of  fermions is defined as $G_{s_1,s_0}(x_1,x_0)\equiv-i\sum_{l} \langle \chi_{l,s_1}(x_1)\chi_{l,s_0}(x_0)\rangle$, where  $s_\alpha \in (+1,-1)$  denote  upper or lower part of the Keldysh contour. The saddle-point Schwinger - Dyson equations read, for this model in the limit 
$N\gg1$, $|x_1-x_0|\gg \frac{T}{J}$ and at $\Phi=0$:
\begin{eqnarray}
\label{SD}
&\Sigma_{s_1,s_0}(x_1,x_0)= s_1 s_0 \left[J^2 G^3_{s_1,s_0}(x_1,x_0)- \Gamma^2 G_{s_1,s_0}(x_1,x_0)\right] \qquad
\Sigma\circ G =-\hat{\mathit{1}}.
\end{eqnarray}
Let us discuss several well-known properties of the original $SYK$ model ($\Gamma=0$). For every solution $\hat{G}$ of Eqs.(\ref{SD}), the function 
\begin{eqnarray}
\label{Gphi}
G^{\phi}_{s_1,s_0}(x_1,x_0)=G_{s_1,s_0}(\phi^{s_1}(x_1),\phi^{s_0}(x_0)) \left[\phi^{s_1\prime}(x_1)\phi^{s_0\prime}(x_0)\right]^\Delta \nonumber \\
\end{eqnarray} 
is also solution;  here $\phi_s$ is   an arbitrary  monotonous function and $\Delta=\frac{1}{4}$; 
we use this generalized notation as it will be useful below for regularization purpose. 
The translation invariant solution has the form: 
$\hat{G}(x_1,x_0)=-i \left(\frac{b}{ J^2}\right)^{\Delta}\hat{g}(x_1-x_0)$ where (see Ref.~\cite{song2017strongly})
\begin{eqnarray}
\hat{g}(x>0)=\left(\frac{1}{4\sinh^2\left(\frac{x}{2}\right)}\right)^\Delta \left(\begin{smallmatrix}
e^{-i\pi \Delta}  && -e^{i\pi \Delta} \\ e^{-i\pi \Delta} &&  -e^{i\pi \Delta}
\end{smallmatrix}\right)
\end{eqnarray}
and $b=4\pi$.
To find Green's function for $x_1<x_0$  one can use the relation $\hat{G}(x_1,x_0)^{T} = -\hat{G}(x_0,x_1)$.  The symmetry group
$SL(2,\mathbb{R})$ of this solution is smaller than the full symmetry group respected by the equations~(\ref{SD}); it
results in the appearance of a soft reparametrization mode and  strong influence of its fluctuations.  To take them into account, we replace each Green function entering the action by $G^{\phi}$ and then integrate over $\phi(x)$ with the following action~\cite{kitaev2018soft,maldacena2016remarks,bagrets2016sachdev,lunkin2018}:
\begin{eqnarray}
\label{action}
&S=S_{SYK}+S_{2}\nonumber \qquad S_{SYK}= -\varepsilon_0 \sum\limits_{s}  \int \{e^{\phi^s(x)},x\} dx \\ &S_2=\frac{ig}{2 \varepsilon_0}\sum\limits_{s_1,s_0}\int dx_1 dx_0 s_1 s_0\left[g^\phi_{s_1,s_0}(x_1,x_0)\right]^2 (1+\Phi^{s_1}(x_1))(1+\Phi^{s_0}(x_0)) \nonumber \\ &g^\phi_{s_1,s_0}(x_1,x_0) =g_{s_1,s_0}(\phi^{s_1}(x_1)-\phi^{s_0}(x_0)) \left[\phi^{s_1\prime}(x_1)\phi^{s_0\prime}(x_0)\right]^\Delta 
\end{eqnarray}
where $\{Y,X\}$ denotes Schwarzian derivate and parameters of the action are given by
\begin{equation}
\varepsilon_0 = \frac{2\pi \gamma}{\beta J}\, ; \quad g= \frac{N\sqrt{b}\gamma\Gamma^2}{2J^2}\, ;
\quad \gamma=\alpha_S N\, ; \quad \alpha_s\approx 0.05.
\label{notations}
\end{equation}
The action (\ref{action}) describes general problem with non-zero $\Gamma$ and parametric modulation $\Phi(x)$,
The magnitudes of $\Phi^{(+,-)}(x) = A\cos(\Omega_P x) + f^{+,-}(x)$  at the upper and the lower branches 
of the Keldysh contour can be different, we denote their difference as "quantum component" $f^+(x)-f^-(x)= f^q(x)$
which is the source field useful for the calculation of  susceptibility; classical component is defined as
$f^{cl} = \frac12 (f^+(x)+f^-(x)) $.

The rate of energy dissipation $W(\omega) a_\omega^2$ due to the presence of oscillating field $f^{cl}(t)=a_\omega\cos(\omega t)$
can be calculated with the help of this action in the following way:
\begin{eqnarray}
&W(\Omega)=\frac{\Omega}{2}\Im \chi(\Omega)\,; \qquad  \chi(\Omega)=-\frac{i}{2}\frac{\delta^2 Z_{\Phi}}{\delta f^{q}_{\Omega}\delta f^{cl}_{\Omega}}
\qquad Z_{\Phi}=\int \mathcal{D}\phi e^{iS} 
\label{eq:def_W}
\end{eqnarray}
In the previous work~\cite{lunkin2020perturbed} we have shown that at $ T \ll \Gamma $ fluctuations of the soft mode are suppressed, while modification of the saddle point solution is small as long as $ T \gg \Gamma^2/J$.  In a broad
intermediate range $\Gamma^2/J \ll T \ll \Gamma$  considered  in the following calculation, we can assume that 
$\phi^s(x)=x+u^s(x)$ where $u^{s\prime}(x)\ll1$. 
Thus  we can work in the Gaussian approximation over $u(x)$ and the action (\ref{action}) take the following form:
\begin{eqnarray}
& S=S_{SYK}+S_{\tilde{\Phi}}\qquad  \qquad S_{\tilde{\Phi}} =S^{(0)}_{\tilde{\Phi}}+S^{(1)}_{\tilde{\Phi}}+S^{(2)}_{\tilde{\Phi}} \nonumber \\
& S_{SYK}=\frac{1}{2}\int \frac{d\Omega}{2\pi} \left[\hat{\mathcal{G}}_0(\Omega)\right]^{-1}_{s_1,s_2}u^{s_1}_{-\Omega} u^{s_2}_{\Omega} \qquad \qquad S^{(0)}_{\tilde{\Phi}} = i\frac{g}{2\varepsilon_0}\int \frac{d\Omega}{2\pi}L^{(0)}_{s_1,s_2}(\Omega)\tilde{\Phi}^{s_1}_{-\Omega}\tilde{\Phi}^{s_2}_{\Omega} \nonumber \\
& S^{(1)}_{\tilde{\Phi}} = i\frac{g}{2\varepsilon_0}\int \frac{d\Omega_0d\Omega_1}{(2\pi)^2}L^{(1)}_{s_1,s_2,s_3}(\Omega_0,\Omega_1)u^{s_1}_{\Omega_0}\tilde{\Phi}^{s_2}_{-\Omega_1-\frac{\Omega_0}{2}}\tilde{\Phi}^{s_3}_{\Omega_1-\frac{\Omega_0}{2}} \nonumber \\
& S^{(2)}_{\tilde{\Phi}}=i\frac{g}{2\varepsilon_0}\int \frac{d\Omega_0d\Omega_1d\Omega_2}{(2\pi)^3}L^{(2)}_{s_1,s_2,s_3,s_4}(\Omega_0,\Omega_1,\Omega_2)u^{s_1}_{\Omega_0}u^{s_2}_{\Omega_1}\tilde{\Phi}^{s_3}_{-\Omega_2-\frac{\Omega_0+\Omega_1}{2}}\tilde{\Phi}^{s_4}_{\Omega_2-\frac{\Omega_0+\Omega_1}{2}} 
\label{eq: initial_soft_action}
\end{eqnarray} 
Here we have introduced new notations $\tilde{\Phi}^{\pm}(x)=1+\Phi^{\pm}(x)$ and $\hat{L}^{(i)}$ are tensors obtained  from the original action (see Appendices \ref{App_1},\ref{App_2}).
This action is Gaussian in terms of fluctuations of $u(x)$ so we can easily do  the  functional integral in Eq.(\ref{eq:def_W}). The propagator of $u(x)$ in this model is determined  by the quadratic over $u(x)$
form in following action:
\begin{eqnarray}
\label{S2u}
S_0=S_{SYK}+S^{(2)}_{\tilde{\Phi}_0}
\end{eqnarray}
where $S^{(2)}_{\tilde{\Phi}_0} $ denotes the term $S^{(2)}_{\tilde\Phi}$ from Eq.(\ref{eq: initial_soft_action}), evaluated at
$\tilde{\Phi}^{\pm}_0=\tilde{\Phi}^{\pm}|_{f=0}=1+A\cos{\Omega_P x} $.
It is useful to understand the functional $Z_{\Phi}$  defined in Eq.(\ref{eq:def_W}) as 
$Z_{\Phi}=\langle e^{i\left(S-S_0\right)} \rangle_0$ where $\langle \ldots \rangle_0$ means average with action $S_0$;
the difference $S - S_0$ contains source terms needed to calculate susceptibility.
Let us analyze the propagator corresponding to this action.

\section{Fluctuation propagator without pumping.}

The action of the soft mode was calculated in Apendices \ref{App_1},\ref{App_2},\ref{App_3} and  has the form
\begin{eqnarray}
& S_{0}= \int \frac{d\Omega}{4\pi}\hat{u}_{-\Omega}^T \left(\begin{smallmatrix}
0 && \left[\mathcal{G}^A(\Omega)\right]^{-1} \\ \left[\mathcal{G}^R(\Omega)\right]^{-1} && -\mathcal{G}^K(\Omega) \left[\mathcal{G}^A(\Omega)\right]^{-1}\left[\mathcal{G}^R(\Omega)\right]^{-1}
\end{smallmatrix} \right) \hat{u}_{\Omega}
\label{eq:quadratic action}
\nonumber \\
\end{eqnarray}
where
\begin{eqnarray}
&\left[\mathcal{G}^{R}(\Omega)\right]^{-1}=\Omega^2 \left(\varepsilon_0 \left(\Omega^2+1\right)-\frac{g}{2\varepsilon_0}
\psi(-\Omega)\right)\,;
\nonumber \\
& -\mathcal{G}^K(\Omega) \left[\mathcal{G}^A(\Omega)\right]^{-1}\left[\mathcal{G}^R(\Omega)\right]^{-1}
=i\frac{g}{2\varepsilon_0}\pi\Omega^2\, ;
\nonumber \\
&\psi(\Omega)=\Psi\left(\frac{1}{2}+i\Omega\right)-\Psi\left(-\frac{1}{2}\right)\quad \Psi(z)=\partial_z \ln \Gamma(z) \nonumber \\
\label{eq:quadratic action2}
\end{eqnarray}
The action (\ref{eq:quadratic action})  is non-local,
thus $[\mathcal{G}^R]^{-1}$  has  non-zero imaginary part;  as a result, the distribution function $F(\Omega)$ can be determined  by the standard  relation $\mathcal{G}^K(\Omega)\equiv F(\Omega)\left(\mathcal{G}^R(\Omega)-\mathcal{G}^A(\Omega)\right)$. 
We find then  $F(\Omega)=\coth(\pi \Omega)$, corresponding to the equilirium bosonic distribution in the absence of pumping.

For $g\gg\varepsilon_0^2 \Leftrightarrow T\ll \Gamma $ the bosonic Green function $\mathcal{G}(\Omega)$ corresponding to the action 
(\ref{eq:quadratic action},\ref{eq:quadratic action2}) demonstrates resonant behavior with the frequencies $\pm \Omega_R$ and the resonance width $\Omega_W$. These parameters are given by:
\begin{eqnarray}
& \Omega_R^2=\frac{g}{2\varepsilon^2_0}\ln(\Omega_R)\, ; \qquad \frac{\Omega_W}{\Omega_R}\approx\frac{\pi}{4\ln(\Omega_R)} \qquad \frac{g}{2\varepsilon^2_0}=\frac{\sqrt{b}}{\alpha_S\left(4\pi \right)^2}\left(\frac{\Gamma}{T}\right)^2
\end{eqnarray}
The frequency  $\Omega_R$ corresponds to the lowest level-spacing in the polaron problem studied in 
Ref.~\cite{lunkin2020perturbed}, in the  case of  many polaron levels (corresponding to large $\kappa$ parameter  defined in 
Ref.~\cite{lunkin2020perturbed}). It is interesting to note that parameters of this resonance peak are detemined  by 
the ratio  $\frac{\Gamma}{T}$ only, while the largest energy scale of the problem $J$ does not enter here. 
Below we will be interested in the low-$T$ region, $T \ll \Gamma$, where deviations from pure SYK model
are most substantial. Note that  fermionic Green function $G(\epsilon)$ is only weakly (by factors $\propto 1/N$) modified 
in the same parameter range, being close to the "conformal limit" solution, while bosonic Green function $\mathcal{G}(\Omega)$ strongly depends on $\Gamma$ and $T$.

Well-defined resonance behaviour with "quality factor" $\ln(\Omega_R) \sim \ln\frac{\Gamma}{T} \gg 1$ is rather surprising to find
in our problem which lacks any apparent energy scale determining the frequency of this resonance. While $\Gamma$ itself is just
the width of the "single-particle  band" that occurs due to $\Gamma_{ij}$ matrix elements, the frequency $\omega_R$
and width $\omega_W$ of the resonance are given, in physical units by
\begin{eqnarray}
\label{freq-res}
\omega_R=2\pi T\Omega_R = \frac{\Gamma}{4\alpha_s\sqrt{\pi}}\ln^{1/2}\frac{\Gamma}{T}
\\ 
\omega_W = 2\pi T\Omega_W = \frac{\sqrt{\pi}}{16\alpha_s}\frac{\Gamma}{\ln^{1/2}\frac{\Gamma}{T}}
\label{freq-W}
\end{eqnarray}  
They are larger (correspondingly, smaller) than $\Gamma$ by the factor
$\sqrt{\ln\frac{\Gamma}{T}} $, coming  from  strong interaction between fermions.  Note that this result is specific
for SYK$_4$ model of interaction.  For any SYK$_{2k}$ model with $k>2$ similar derivation would lead to some expressions of the
type of Eqs.(\ref{eq:quadratic action},\ref{eq:quadratic action2}) with the function $\psi(\Omega)$ replaced by some power-law
function $\tilde\psi(\Omega) \sim \Omega^\alpha $ where $\alpha={4}/{k}-1$; as a result the ratio $\frac{\Omega_W}{\Omega_R}\propto \frac{\Im \psi(\Omega_0)}{\Re \psi(\Omega_0)}\sim O(\alpha)$ would be of the order of unity. The SYK$_4$ model is special since it leads, effectively,
to $\alpha=0$.
The resonant behaviour  leads to a substantial change of the dissipation rate compared to the original SYK$_4$ model.
In addition, the system properties can be seriously modified  by applying \textit{ac} pumping with frequency
$\omega_P\approx \omega_R$.

\section{Dissipation rate in the linear regime.}

To find susceptibility $\chi(\Omega)$ we need to calculate the average $[S^{(1)}]^2$, while the term $S^{(2)}$ can be neglected. The result reads:
\begin{eqnarray}
\chi(\Omega)=
\frac{1}{2}\frac{\delta^2 }{\delta f^{q}_{\Omega}\delta f^{cl}_{\Omega}}\left(S^{(0)}_{\tilde{\Phi}}+\frac{i}{2}\langle [S^{(1)}]^2  \rangle\right)
\label{chi-1}
\end{eqnarray}
Let us first discuss the case of pure SYK model (that is, $\Gamma=0$); then  susceptibility is determined by the first term of (\ref{chi-1}) only:
\begin{equation}
\label{chi-2}
\chi_{SYK}(\Omega)=\frac{2g}{ \varepsilon_0}\left(\ln\left(\frac{\beta J}{2\pi}\right)-\psi(-\Omega)\right) 
\end{equation} 
To get dissipative part of physical susceptibility, we need also to multiply $\Im\chi_{SYK}(\Omega)$ by the factor 
$2\pi T = \omega/\Omega$; the result is (we used also relations (\ref{notations})):
\begin{equation}
\Im\chi_{SYK}(\omega)= 2\pi T \frac{\pi g}{ \varepsilon_0} \tanh\frac{\omega}{2T} =
\pi^{3/2}\frac{N\Gamma^2}{J}  \tanh\frac{\omega}{2T} 
\label{chi-3}
\end{equation}
The logarithmically large term in Eq.(\ref{chi-2}) comes due a formal divergence in the integral over $d x_1$ in the term $S_2$
in the action (\ref{action}). This divergence is present since $g(x_1,x_0)\propto |x_1-x_0|^{-1}$  at $x_1\rightarrow x_0$.  We need to cut-off this integral at the scale $[\Delta x]_{min} \sim \frac{J}{T}$ since at lower $\Delta x$ our long-wavelength expression for the action is not valid.
Below we will omit this term from the expressions for susceptibility, since it is just real constant independent on $\Omega$.
The result (\ref{chi-3}) is known  in the theory of non-Fermi liquid~\cite{parcollet1999non,varma1989phenomenology}; it is also related to linear dependence of  resistance on  temperature in the model of Ref.~\cite{song2017strongly}. 

Inclusion of quadratic terms $\sim \Gamma_{ij}$ into the Hamiltonian leads to strong modification of the $u(x)$ fluctuation propagator, and thus to a considerable change in susceptibility.
Detailed calculation of $\frac{i}{2}\langle [S^{(1)}]^2$  can be found in the appendix \ref{App_4}. 
After some algebra, one can find
susceptibility in the following form:
\begin{eqnarray}
\chi(\Omega)=  -\frac{2g}{ \varepsilon_0}\psi(-\Omega) \left[1+\frac{g}{ 2\varepsilon_0} 
\Omega^2\mathcal{G}^{R}_{\Omega} \psi(-\Omega)\right]
\label{eq: general expression for susceptibility } 
\end{eqnarray}
 This formula can be understood in the following way. Let us consider a representation for the  $S_2$ from Eq.(\ref{action}) using Keldysh contour $\mathcal{C}$: 
\begin{eqnarray}
S_2 = i \frac{g}{2\varepsilon_0}\int\limits_{\mathcal{C}}dx_1 dx_0 \left[g^{\phi}(x_1,x_0)\right]^2 (1+\Phi(x_1))(1+\Phi(x_0)).
\end{eqnarray} 
Now we  consider arbitrary re-parametrization of this contour $x \mapsto \tilde{x}(x)$. We can rewrite this part of the action using the re-parametrization as follows: 
\begin{eqnarray}
S_2 = i \frac{g}{2\varepsilon_0}\int\limits_{\mathcal{C}}d\tilde{x}_1 d\tilde{x}_0 \left[g^{\phi}(\tilde{x}_1,\tilde{x}_0)\right]^2 (1+\Phi(\tilde{x}_1))(1+\Phi(\tilde{x}_0)) \left[\frac{\partial \tilde{x}_1}{\partial x_1}\frac{\partial \tilde{x}_0}{\partial x_0}\right]^{2\Delta-1}.
\end{eqnarray}
Now we can fix the choice of re-parametrization as the solution of the following  equation: $(1+\Phi(\tilde{x}))\left[\frac{\partial \tilde{x}}{\partial x}\right]^{2\Delta-1}=1$. It will simplify the action and leads to the in the form: \newline $ S_2 = i \frac{g}{2\varepsilon_0}\int\limits_{\mathcal{C}}d\tilde{x}_1 d\tilde{x}_0 \left[g^{\phi}(\tilde{x}_1,\tilde{x}_0)\right]^2$. As a result we will find the action written in the term of the function $\phi(\tilde{x})$. In the linear approximation we can write: \begin{eqnarray}
\phi(\tilde{x})\approx\tilde{x}+u(x)\approx x+u(x)+\frac{1}{1-2\Delta}\int_{-\infty}^{x} \Phi(x^\prime) dx^\prime.
\end{eqnarray}
 It means that we can write the quadratic  action and, as a result, the effective action for the sources  in the form:
\begin{eqnarray}
&S_{SYK}=\frac{1}{2}\int \frac{d\Omega}{2\pi} \left\{ u^{\dagger}_{\Omega}\left[\hat{\mathcal{G}}_0(\Omega)\right]^{-1}u_{\Omega}-\left(u_{\Omega}+\frac{2i}{\Omega} \Phi_{\Omega}\right)^{\dagger} \hat{\Sigma}(\Omega) \left(u_{\Omega}+\frac{2i}{\Omega} \Phi_{\Omega}\right)  \right\} \nonumber \\ 
&S_{eff} = \frac{1}{2}\int \frac{d\Omega}{2\pi} \frac{4}{\Omega^2} \Phi_{\Omega}^{\dagger}  \left\{\hat{\Sigma}(\Omega)+\hat{\Sigma}(\Omega) \hat{G}(\Omega) \hat{\Sigma}(\Omega)  \right\} \Phi_{\Omega}  \nonumber \\
&\hat{\Sigma}(\Omega)=\frac{g}{2\varepsilon_0}\Omega^2\left(\begin{smallmatrix}
0 & \psi(\Omega) \\  \psi(-\Omega) & -i\pi
\end{smallmatrix}\right)
\end{eqnarray}
The special structure  $\Sigma + \Sigma G \Sigma $ in the action $S_{eff}$ leads to the relation
(\ref{eq: general expression for susceptibility }) between observable susceptibility $\chi(\Omega)$ and bosonic Green function
$G(\Omega)$. We note that the above consideration is not applicable to a ultraviolet-singular part of the susceptibility $\ln(\beta J/2\pi)$,
but it is not important for our analysis since no contribution to $\Im\chi(\Omega)$ comes from the ultraviolet.


The effect of quadratic perturbations is best represented  by the imaginary part of susceptibility: 
\begin{eqnarray}
\Im \chi(\Omega)= \frac{\left(\Omega^2+1\right)^2 \Im\chi_{SYK}(\Omega)}{ \left(\Omega^2+1-\frac{g}{2\varepsilon^2_0}\Re\psi(-\Omega)\right)^2+\left[\frac{g}{2\varepsilon^2_0}\Im\psi(-\Omega)\right]^2} \nonumber \\
\label{chi-4}
\end{eqnarray}
Under the condition $\Gamma\gg T \Leftrightarrow g\gg\varepsilon_0^2$, formula (\ref{chi-4}) demonstrates resonance peak
at $\Omega \approx \Omega_R$.
At high frequencies $\Omega \gg \Omega_R$ new result (\ref{chi-4}) reduces to pure SYK one. Near resonance, at
$\Omega\approx \Omega_R$, dissipation in our model is enhanced by the factor 
\begin{equation}
\frac{\Im \chi (\Omega)}{\Im\chi_{SYK}(\Omega)}=\left(\frac{\Omega_R }{ 2 \Omega_W}\right)^2 = 
\frac4{\pi^2}\ln^2\Omega_R \gg 1
\label{chi-peak}
\end{equation}
\begin{figure}
	\centering
	\includegraphics[width=0.9\linewidth]{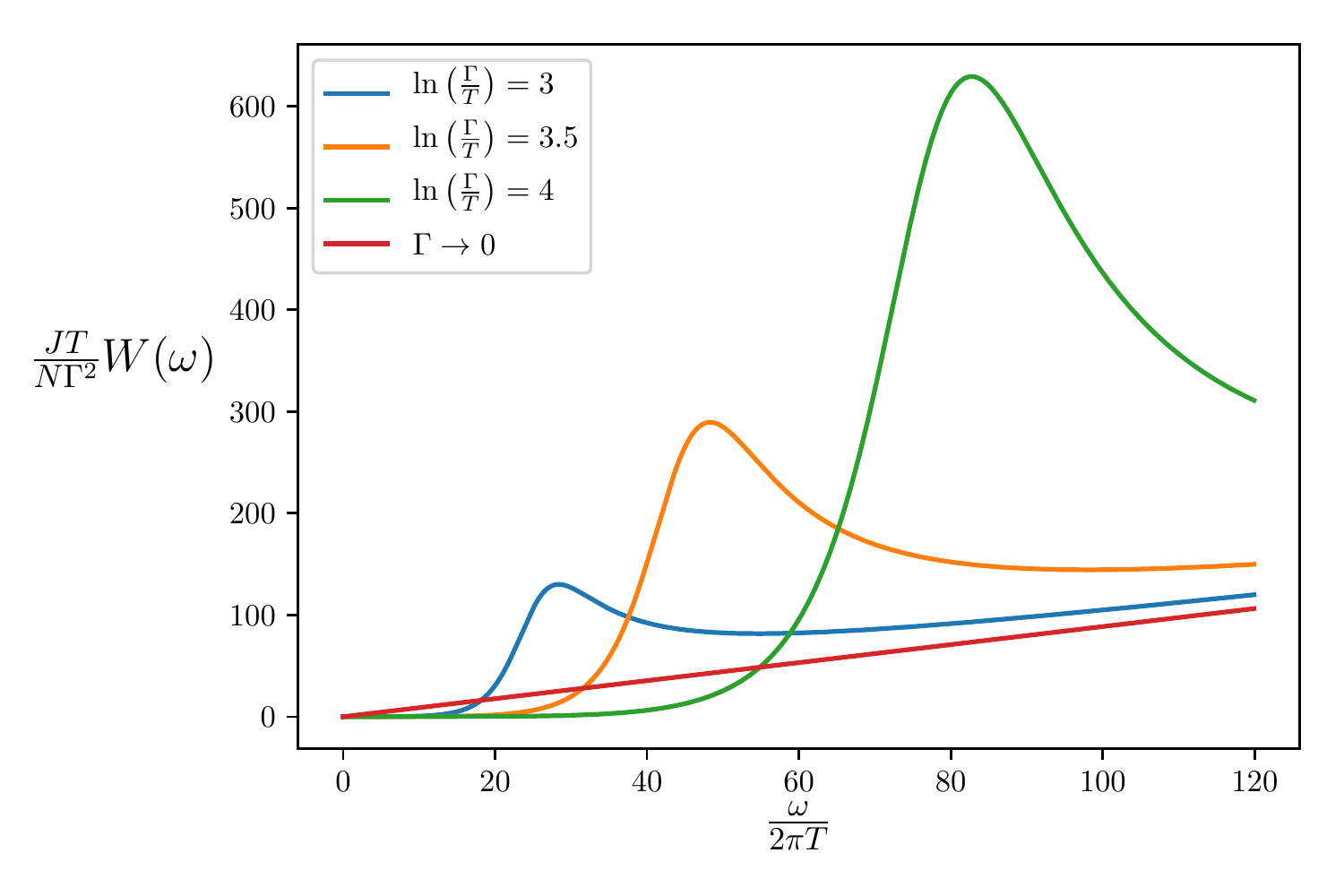}
	\caption{Dissipation rate at large $\Gamma/T$ and at $\Gamma\to 0$; resonant enhancement is seen at
		$\omega \approx \omega_R$. Note that low-frequency part of the figure with strong suppression of $W(\omega)$ 
		corespond to the contribution of the soft mode only; full magnitude of the rate should include here also the
		contribution from Eq.(\ref{chi-M}).}
	\label{fig:graph11}
\end{figure}
Frequency-dependence of  dissipation rate $W(\omega)$  corresponding to Eq.(\ref{chi-4}) is shown in Fig.(\ref{fig:graph11})
for several large values of $\Gamma/T$ ratio.

At low frequencies $\Omega\ll\Omega_W$ dissipation given by Eq.(\ref{chi-4}) is strongly suppressed w.r.t. pure SYK model:
\begin{equation}
\frac{\Im \chi(\Omega)}{\Im\chi_{SYK}(\Omega)}=\left(\frac{\Omega^2+1 }{ 2\Omega_R \Omega_W|\psi(-\Omega)|}\right)^2
\sim \left(\frac{(2\pi T)^2+\omega^2}{\Gamma^2}\right)^2
\label{chi-low}
\end{equation}
Applicability of the result (\ref{chi-low}) is limited since it is obtained for the temperature region $ T \ll \Gamma$
without account for modification of the saddle-point solution which is known to lead to Fermi-liquid saddle point at
$T \leq T_{FL} = \Gamma^2/J$. 
In other terms, in the above calculation we did not account for the hard modes which control the crossover to the FL state.
Such a calculation in provided in the appendix \ref{App_5}, and it results in the following additional term to dissipative part of susceptibility written below for $\omega \ll T$:
\begin{equation}
\Im\chi_{\mathrm{add}}(\omega) =  \mathcal{C}\, \omega N \mathcal{M}^2  =  \frac{\mathcal{C}}{(4\pi)^2}\, \omega N 
\left(\frac{T_{FL}}{T}\right)^2
\label{chi-M}
\end{equation}
where numerical coefficient $\mathcal{C}\approx 234$. At $T \sim T_{FL}$ the contribution (\ref{chi-M}) is of the same order
of magnitude as the Fermi-liquid result $\Im\chi_{FL}(\omega) \sim N\omega$.
The contribution (\ref{chi-M}) becomes comparable to the soft-mode contribution (\ref{chi-low}) at $T \approx T_{cr} = \Gamma (\Gamma/J)^{1/5}$.
Thus we see that Eq.(\ref{chi-low}) is valid in the narrow range $\Gamma (\Gamma/J)^{1/5} < T < \Gamma$.

\section{Nonlinear pumping effects.}
 We consider now the behavior of our system at non-equilibrium conditions, under the external \textit{ac} pumping with frequency $\omega_P \approx \omega_R$ and amplitude $A$,
thus $\Phi(t) = A\cos(\omega_P t) + f(t)$.  The difference $|\omega_P -\omega_R|$ is irrelevant as long as it is
much smaller than resonance width $\omega_W$, and below we will assume $\omega_P=\omega_R$.
We are interested here in the modification of  low-frequency  response at $\omega \ll \omega_W$ due to such a pumping.

Important remark is in order: pumping leads to energy absorption by our system, thus for the stationary distribution to
exist, some kind of coupling to external bath is necessary. Here we prefer to employ another approach: we consider
finite-time pumping during timescale $t_{\mathrm{pump}}$ choosen in such a way that the total energy absorbed
in our system $E_{\mathrm{abs}} = W(\omega) A^2 t_{\mathrm{pump}}$ is small enough, so the increase $\Delta T$ of its temperature is  relatively small,
$\Delta T \ll T$.  In this respect our approach is  very different from  the one developed in Ref.~\cite{Driven} for the $N=\infty$ limit of
the combined SYK$_4$+SYK$_2$  system: they studied the limit of strong energy pumping.
Then we can study quasi-stationary response of our system
at frequencies $\omega$ restricted  by the condition
$\omega \gg 1/t_{\mathrm{pump}}$. Increase of the temperature due to external pumping source is $\Delta T = E_{\mathrm{abs}}/C(T)$ where $C(T)$ is the heat capacity.
The major contribution to $C(T)$ in the interesting range $T \ll \Gamma$ comes from quadratic terms in the action. Heat capacity $C(T)$ can be calculated in the saddle-point approximation, see appendices \ref{App_2_3} and \ref{App_5};
the corresponding calculation of the entropy was performed
in~\cite{song2017strongly}, but it was left unnoticed that dominant at $ T < \Gamma$ contribution to $C(T)$ grows as  $\sqrt{4\pi} N\Gamma^2/JT$
before reaching the maximum at $T\approx \Gamma^2/J$
$C(T) \approx \sqrt{4\pi} N\Gamma^2/JT$. In result, we find sequence of inequlities for magnitudes of $A$ and $\omega$:
\begin{equation}
\omega_R A^2 \ln^2\frac{\omega_R}{T} \ll \frac1{t_{\mathrm{pump}}} \ll \omega
\label{restr}
\end{equation}

Nonzero pumping amplitude affects low-frequency susceptibility in two ways: first, it changes the action of the soft
mode; second, it creates a correction to the term $\frac{i}{2}\langle [S^{(1)}]^2  \rangle $ since 
$S^{(1)}$ is of the second order in $\tilde{\Phi}$. The second contribution occurs to be small at $\omega \ll \omega_W$,
as it is shown in the appendix \ref{App_4_4}, so we neglect it.

Pumping-induced corrections to the action of the soft mode comes in two ways. First contribution which we call "direct" one,
is due to the terms $\sim A^2$ and contains terms like $u_{-\Omega}u_{\Omega}$ in the action. Second contribution is "indirect"
in the sense that it is $\propto A$ and it contains mixture of high- and low-frequency harmonics, 
$u_{-\Omega}u_{\Omega+\Omega_R}$. After  Gaussian integration over fast modes $u_{\Omega_f}$ with $\Omega_f \approx \Omega_R$, these terms also produce contribution to the action of slow soft modes. The combination of direct and indirect terms
leads (for details see appendix \ref{App_3}) to the additional action of the soft mode:

\begin{eqnarray}
\label{deltaA}
\delta_A S_{soft}=\pi \Omega_P\frac{ig}{2 \varepsilon_0}\left(\frac{A}{2}\right)^2\int \frac{d\Omega}{(2\pi)} \hat{u}_{-\Omega}^T \left(\begin{smallmatrix}
0 && -\Omega \\ \Omega && 4 \Omega_P
\end{smallmatrix}\right)\hat{u}_{\Omega} \nonumber\\
\end{eqnarray}
Off-diagonal terms in the $2\times 2$ matrix in Eq.(\ref{deltaA}) are proportional to $\Omega$, contrary to the analogous terms
in the original action $S_{soft}$, see Eq.(\ref{eq:quadratic action}), which starts from terms $\propto\Omega^2$. These
linear in $\Omega$ terms indicates the appearence of \textit{friction}  due to non-equilibrium nature of the system under
pumping.
Apart from modification of quadratic in $u(x)$ terms, pumping leads also to the appearence of the singular contribution 
$\propto A^2 t_{\mathrm{pump}}$ to the average value $\langle u(x) \rangle$ which reflects the raise of temperature 
$T \to T+\Delta T$ due to pumping, see appendix \ref{App_5}. 
In the harmonic approximation over $u(x)$ we used, the average  $\langle u(x) \rangle$ does not change fluctuation propagator,
however it will be affected once non-linear in $u(x)$ terms will be taken into account.
The condition (\ref{restr}) ensures that $\Delta T\ll T$ and this effect is small.

Combining two contributions to the action, Eqs.(\ref{eq:quadratic action}) and (\ref{deltaA}), we obtain full action
for low-frequency soft mode fluctuations. The corresponding retarded Green function is
\begin{eqnarray} 
&\left[\mathcal{G}^{R}(\Omega)\right]^{-1}=\Omega^2 \left(\varepsilon_0 \left(\Omega^2+1\right)-\frac{g}{2\varepsilon_0}\psi(-\Omega)\right)+\nonumber \\&\frac{i\pi \,g\, \Omega_P  }{ \varepsilon_0}\Omega\left(\frac{A}{2}\right)^2 .
\label{GrA} 
\end{eqnarray}

Now we can use Eq.(\ref{eq: general expression for susceptibility }) to find susceptibility $\chi(\omega)$ with respect to
\textit{ac} probe field $f_\Omega$  in the  range $\frac{1}{T t_{pump}}\ll\Omega \ll \Omega_R$, see also Eq.(\ref{restr}):
\begin{eqnarray}
\label{chiA}
\Im{\chi}(\Omega)=\frac{\pi g}{\varepsilon_0} \left(\frac{\Omega_R }{\Omega}A^2+\frac{(\Omega^2+1)^2}{|\frac{g}{2\varepsilon_0^2}\psi(-\Omega)|^2}\tanh(\pi \Omega)\right) \nonumber \\
\end{eqnarray}

\begin{figure}
	\centering
	\includegraphics[width=0.95\linewidth]{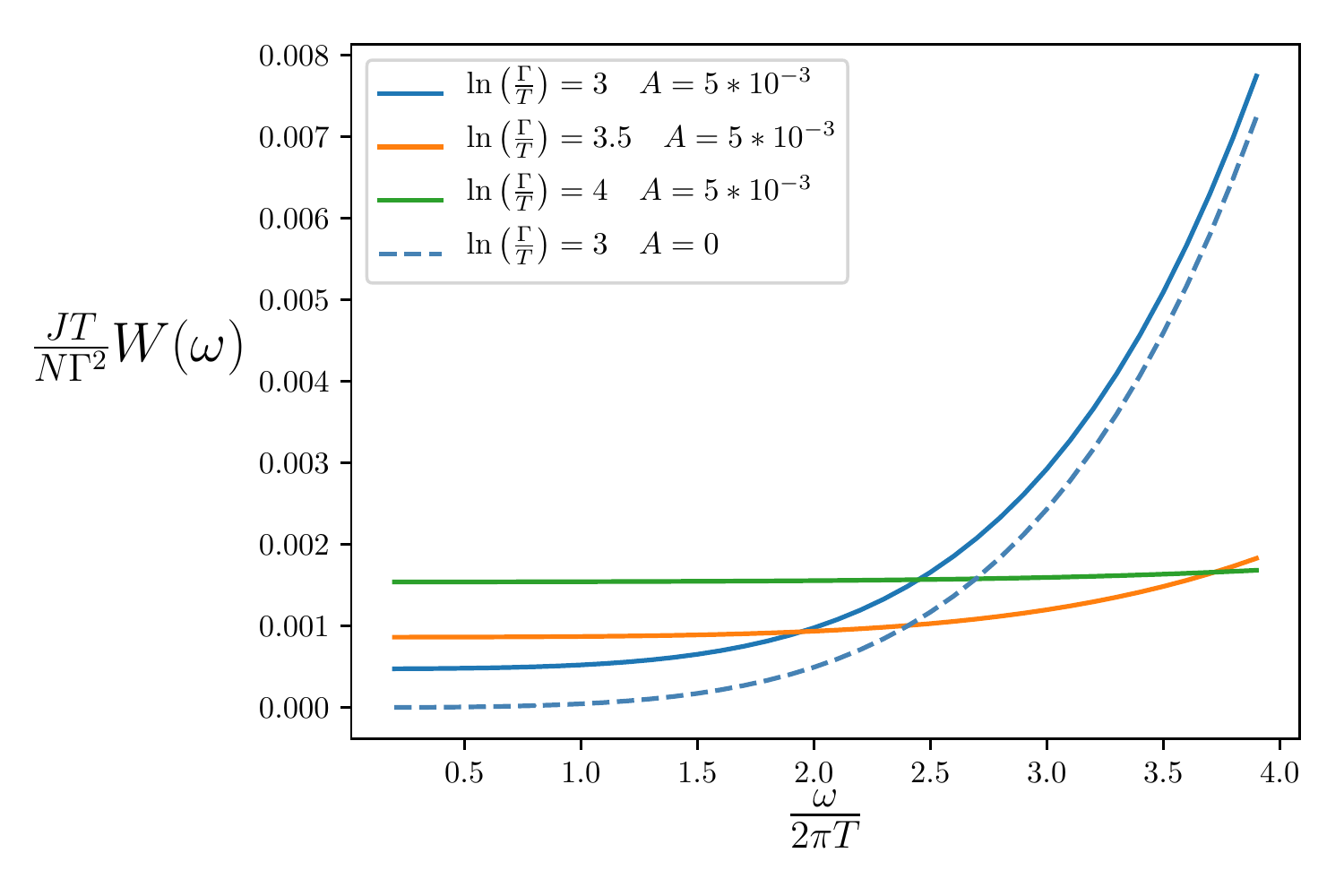}
	\caption{Dissipation rate at small frequencies $\omega \ll \omega_R$ in the presence of pumping with amplitude
		$A=5\cdot 10^{-3}$.
		Very weak $\omega$-dependence at intermediate frequencies correspond to the "dry friction domain",  Eq.(\ref{chiA-1}).
	}
	\label{fig:graph31}
\end{figure}

Equation (\ref{chiA}) demonstrates the presence of the typical scale $I(\Omega)=\frac{\Omega(\Omega^2+1)}{\Omega_R^2\Omega_W}$ 
for the pumping amplitude  $A$; the response frequency $\Omega \ll \Omega_W$.
At $A \ll I(\Omega)$ pumping is weak and does not modify susceptibility and dissipation rate.  
In the range $I(\Omega) \ll A $ one finds 
$\Im\chi^A_1(\Omega)\approx \pi A^2\frac{g}{ \varepsilon_0} \frac{\Omega_R }{\Omega} $ and dissipation rate
\begin{eqnarray}
W(\omega)= \frac{\omega}{2}\Im\chi(\omega) = \frac{A^2}{2} N \pi^{3/2} \omega_R \frac{\Gamma^2}{ J}
\label{chiA-1}
\end{eqnarray}
which is somewhat unusual:  $W(\omega)$ does not depend on $\omega$ and $T$, which reminds a "dry friction" phenomenon.
Fig. (\ref{fig:graph31}) represents $W(\omega)$ behavior as it follows from Eq.(\ref{chiA}), for the  magnitudes of pumping $A=5\cdot 10^{-3}$ and several small ratios ${T}/{\Gamma} \ll 1$.

\section{Conclusion}
We have studied energy absorption in a driven system of strongly correlated  Mayorana fermions of mixed
SYK$_4$ + SYK$_2$ type, with relatively weak quadratic part of the Hamiltonian, $\Gamma \ll J$, at the temperatures
$T \gg T_{FL}=\Gamma^2/J$. Our major results refer to the region $T \leq \Gamma$ where new type of unviersal behaviour
was found: fluctuations of the soft mode are characterized by the resonant behaviour with frequency $\omega_R$ and
width $\omega_W$ given by Eqs.(\ref{freq-res},\ref{freq-W}).
Quality factor of this resonance is $Q = \frac4{\pi} \ln\frac{\Gamma}{T}$. Suprisingly,
both $\omega_R$ and $\omega_W$ do not depend on the largest interaction energy scale $J \gg \Gamma$.
The frequency $\omega_R$ directly corresponds to the polaron bound-state energy found in Ref.~\cite{lunkin2020perturbed} 
via Matsubara-time approach. Although Fermionic Green function itself is accurately described by the conformal saddle-point solution,
physical properties appears to be sensitive to the bosonic collective mode that becomes instrumental at $T < \Gamma$.

Physical significance of the polaron soft mode is demonstrated via the results we obtained for the energy
dissipation due to parametric modulation of the quadratic part of the Hamiltonian.  
For near-resonance modulation  frequencies $\omega \approx \omega_R$  it  is found to be enhanced by the factor 
$\frac14 Q^2$  w.r.t. pure SYK$_4$ model, see Eqs.(\ref{chi-4},\ref{chi-peak}).
On the contrary, at lower frequencies $\omega \ll \omega_R$ dissipation rate is suppressed, see Eqs.(\ref{chi-low},\ref{chi-M}).

Pumping the system with \textit{ac} modulation of finite amplitude $A$ at the resonant frequency  $\omega_R$
leads to a non-equilibrium state those response to a linear low-$\omega$ perturbation differs considerably from the case of
$A=0$. First, the pumping leads to a "friction term" in the action of the soft mode, see Eqs.(\ref{deltaA}),(\ref{GrA}).
Secondly, the power  $W(\omega)$ dissipated due to \textit{ac} modulation is nearly $\omega$-independent in the range
$ I(\omega) \ll A^2$, see Fig.\ref{fig:graph31}.

In spatially extended systems the polaron soft mode is expected to be related to  the energy transport.
Rich behaviour of its propagator $\mathcal{G}(\omega)$  as function of $\Gamma$, $T$ and frequency indicates
that transport properties of extended SYK-based models may occur to be more diverse than it seems to follow from
the saddle-point analysis~\cite{song2017strongly}.

\section*{Acknowledgements}
We are grateful to A. Yu. Kitaev for numerous useful discussions and to K. S. Tikhonov for 
important comments.

\paragraph{Funding information}
 Research of A.V.L. was partially supported by the Basis Foundation,  by 
the Basic research program of the HSE and by the RFBR grant \# 20-32-90057.

\begin{appendix}

\section{The effective action}
\label{App_1}
The action of the model is present the main text and has the following form:
\begin{eqnarray}
&S=S_{SYK}+S_{2}\nonumber \quad S_{SYK}= -\varepsilon_0 \sum\limits_{s}  \int \{e^{\phi^s(x)},x\} dx\\ &S_2=\frac{ig}{2 \varepsilon_0}\sum\limits_{s_1,s_0}\int dx_1 dx_0 s_1 s_0\left[g^\phi_{s_1,s_0}(x_1,x_0)\right]^2 (1+\Phi^{s_1}(x_1))(1+\Phi^{s_0}(x_0)) \nonumber \\ &g^\phi_{s_1,s_0}(x_1,x_0) =g_{s_1,s_0}(\phi^{s_1}(x_1)-\phi^{s_0}(x_0)) \left[\phi^{s_1\prime}(x_1)\phi^{s_0\prime}(x_0)\right]^\Delta \nonumber \\
&\varepsilon_0 = \frac{2\pi \gamma}{\beta J} \quad g= \frac{N\sqrt{b}\gamma\Gamma^2}{2J^2}\quad \gamma=\alpha_S N\quad \alpha_s\approx 0.05 \nonumber \\
&\hat{g}(x)=\left(\frac{1}{4\sinh^2\left(\frac{x}{2}\right)}\right)^\Delta\left[ \theta(x)\left(\begin{smallmatrix}
e^{-i\pi \Delta}  && -e^{i\pi \Delta} \\ e^{-i\pi \Delta} &&  -e^{i\pi \Delta}
\end{smallmatrix}\right)+ \theta(-x)\left(\begin{smallmatrix}
-e^{-i\pi \Delta}  && -e^{-i\pi \Delta} \\ e^{i\pi \Delta} &&  e^{i\pi \Delta}
\end{smallmatrix}\right)\right]
\end{eqnarray}
Our aim is to calculate susceptibility with respect to the field $\Phi$, which is defined as follows:
\begin{eqnarray}
\chi^R(t-t^\prime)=-\frac{i}{2}\frac{\delta^2 Z_{\Phi}}{\delta \Phi^q(t)\delta \Phi^{cl}(t^\prime)}\quad Z_{\Phi}=\int \mathcal{D}\phi e^{iS}
\end{eqnarray}
We  consider the limit of weak fluctuations, thus the action can be written in the form
\begin{eqnarray}
& S=S_{SYK}+S_{\tilde{\Phi}}\quad S_{\tilde{\Phi}} =S^{(0)}_{\tilde{\Phi}}+S^{(1)}_{\tilde{\Phi}}+S^{(2)}_{\tilde{\Phi}} \quad S_{SYK}=\frac{1}{2}\int \frac{d\Omega}{2\pi} \left[\hat{\mathcal{G}}_0(\Omega)\right]^{-1}_{s_1,s_2}u^{s_1}_{-\Omega} u^{s_2}_{\Omega} \nonumber \\
& S^{(0)}_{\tilde{\Phi}} = i\frac{g}{2\varepsilon_0}\int \frac{d\Omega}{2\pi}L^{(0)}_{s_1,s_2}(\Omega)\tilde{\Phi}^{s_1}_{-\Omega}\tilde{\Phi}^{s_2}_{\Omega} \quad S^{(1)}_{\tilde{\Phi}} = i\frac{g}{2\varepsilon_0}\int \frac{d\Omega_0d\Omega_1}{(2\pi)^2}L^{(1)}_{s_1,s_2,s_3}(\Omega_0,\Omega_1)u^{s_1}_{\Omega_0}\tilde{\Phi}^{s_2}_{-\Omega_1-\frac{\Omega_0}{2}}\tilde{\Phi}^{s_3}_{\Omega_1-\frac{\Omega_0}{2}} \nonumber \\
& S^{(2)}_{\tilde{\Phi}}=i\frac{g}{2\varepsilon_0}\int \frac{d\Omega_0d\Omega_1d\Omega_2}{(2\pi)^3}L^{(2)}_{s_1,s_2,s_3,s_4}(\Omega_0,\Omega_1,\Omega_2)u^{s_1}_{\Omega_0}u^{s_2}_{\Omega_1}\tilde{\Phi}^{s_3}_{-\Omega_2-\frac{\Omega_0+\Omega_1}{2}}\tilde{\Phi}^{s_4}_{\Omega_2-\frac{\Omega_0+\Omega_1}{2}}
\end{eqnarray} 
Here we have introduced $\tilde{\Phi}^{\pm}(x)=1+\Phi^{\pm}(x)$ and $\hat{L}^{(i)}$ are tensors which we will obtain below using Taylor expansion over $S_2$. Using these expressions we will analyze  modified new propagator of soft modes and calculate the dissipation rate  $W(\omega)= \frac{\omega}{2}\Im \chi^R(\omega)$.

\section{The Taylor expansion in powers of $S_2$.}
\label{App_2}
\subsection{Frequency domain}
\label{App_2_1}
We will start our analysis from representation of  $s_1 s_0\left[g_{s_1,s_0}(x)\right]^2$:
\begin{eqnarray}
&s_1 s_0\left[g_{s_1,s_0}(x)\right]^2= \left(\frac{1}{4\sinh^2\left(\frac{x}{2}\right)}\right)^d\left[ \theta(x)\left(\begin{smallmatrix}
e^{-i\pi d}  && -e^{i\pi d} \\ -e^{-i\pi d} &&  e^{i\pi d}
\end{smallmatrix}\right)+ \theta(-x)\left(\begin{smallmatrix}
e^{-i\pi d}  && -e^{-i\pi d} \\ e^{i\pi d} &&  -e^{i\pi d}
\end{smallmatrix}\right)\right]_{s_1,s_0}=\int \frac{d\Omega}{2\pi} e^{-i \Omega x } \hat{L}_{s_1,s_0}(\Omega)\nonumber \\
&\hat{L}(\Omega) =  \left[ K_d(\Omega)\left(\begin{smallmatrix}
e^{-i\pi d}  && -e^{i\pi d} \\ -e^{-i\pi d} &&  e^{i\pi d}
\end{smallmatrix}\right)+ K_d(-\Omega)\left(\begin{smallmatrix}
e^{-i\pi d}  && -e^{-i\pi d} \\ -e^{i\pi d} &&  e^{i\pi d}
\end{smallmatrix}\right)\right] \nonumber \\
&K_d(\Omega) = \int_0^{\infty}\frac{dx}{2\pi} e^{-i \Omega x}  \left(\frac{1}{4\sinh^2\left(\frac{x}{2}\right)}\right)^d = \frac{\Gamma(1-2d)\Gamma(d-i\Omega)}{\Gamma(1-d-i\Omega)}
\end{eqnarray}
Here we have introduced $d=2\Delta \rightarrow \frac{1}{2}-0$ we need this parameter for proper limit as we see that $K_d(\Omega)$ is divergence at $d=\frac{1}{2}$. The matrix $L(\Omega)$ plays the crucial role in our calculation. Using above expression we can write the terms from $S_2$ contained dependence on $u$ where $u$ is defined as $\phi^{\pm}(x)=x+u^{\pm}(x)$:
\begin{eqnarray}
\label{exp1}
&s_1 s_0\left[g^\phi_{s_1,s_0}(x_1,x_0)\right]^2 = s_1 s_0\left[g_{s_1,s_0}(\phi^{s_1}(x_1)-\phi^{s_0}(x_0))\right]^2 \left(\phi^{s_1\prime}(x_1)\phi^{s_0\prime}(x_0)\right)^d = \nonumber \\ &\int \frac{d\Omega}{2\pi} e^{-i \Omega (\phi^{s_1}(x_1)-\phi^{s_0}(x_0)) } \left(\phi^{s_1\prime}(x_1)\phi^{s_0\prime}(x_0)\right)^d\hat{L}_{s_1,s_0}(\Omega) =  \nonumber \\ &\int \frac{d\Omega}{2\pi} e^{-i \Omega (x_1-x_0) } \hat{L}_{s_1,s_0}(\Omega) \left((1+u^{s_1\prime}(x_1))(1+u^{s_0\prime}(x_0))\right)^d e^{-i \Omega (u^{s_1}(x_1)-u^{s_0}(x_0)) }
\end{eqnarray}
The last line in  (\ref{exp1}) is useful to develop the Taylor series over $u(x)$.

\subsection{Three orders of expansion over $u(x)$}
\label{App_2_2}

For further calculations, we consider the following Taylor expansion:
\begin{eqnarray}
&\left((1+u^{s_1\prime}(x_1))(1+u^{s_0\prime}(x_0))\right)^d e^{-i \Omega (u^{s_1}(x_1)-u^{s_0}(x_0)) }\nonumber \\ & = F_{s_1,s_0}^{(0)}(x_1,x_2)+F_{s_1,s_0}^{(1)}(x_1,x_2)+F_{s_1,s_0}^{(2,1)}(x_1,x_2)+F_{s_1,s_0}^{(2,2)}(x_1,x_2) \nonumber \\
&F_{s_1,s_0}^{(0)}(x_1,x_2) = 1 \quad F_{s_1,s_0}^{(1)}(x_1,x_2) = d(u^{s_1\prime}(x_1)+u^{s_0\prime}(x_0))-i \Omega (u^{s_1}(x_1)-u^{s_0}(x_0)) \nonumber \\
& F_{s_1,s_0}^{(2,1)}= \frac{d(d-1)}{2}\left([u^{s_1\prime}(x_1)]^2+[u^{s_0\prime}(x_0)]^2\right) -\frac{\Omega^2}{2}\left( [u^{s_1}(x_1)]^2+[u^{s_0}(x_0)]^2\right)\nonumber \\&-i \Omega d (u^{s_1}(x_1)u^{s_1\prime}(x_1)-u^{s_0}(x_0)u^{s_0\prime}(x_0))\nonumber \\& F_{s_1,s_0}^{(2,2)}= d^2u^{s_1\prime}(x_1)u^{s_0\prime}(x_0) +\Omega^2 u^{s_1}(x_1)u^{s_0}(x_0)-i \Omega d (u^{s_1}(x_1)u^{s_0\prime}(x_0)-u^{s_0}(x_0)u^{s_1\prime}(x_1)) \nonumber \\ \end{eqnarray}
We have divided the contribution proportional to $u^2$ into two  different parts, $F^{(2,1)}$ and  $F^{(2,2)}$. Each term in $F^{(2,1)}$ depends  on single  variable  $u(x_i)$ whereas every term in  $F^{(2,2)}$ depends on both $u(x_0)$ and $u(x_1)$. 
The expressions for $F^{(a)}$ enters the corresponding terms in the action,  $S^{(i)}$:
\begin{eqnarray}
&S^{(a)}_{\tilde{\Phi}}=\frac{ig}{2 \varepsilon_0}\sum\limits_{s_1,s_0}\int dx_1 dx_0 \int \frac{d\Omega}{2\pi} e^{-i \Omega (x_1-x_0) } \hat{L}_{s_1,s_0}(\Omega) F_{s_1,s_0}^{(a)}(x_1,x_2) \tilde{\Phi}^{s_1}(x_1)\tilde{\Phi}^{s_0}(x_0)  
\end{eqnarray}
where second-order term is composed of two parts:
\begin{equation}
S^{(2)}_{\tilde{\Phi}}=S^{(2,1)}_{\tilde{\Phi}}+S^{(2,2)}_{\tilde{\Phi}}
\end{equation}
The terms $S^{(a)}_{\tilde{\Phi}}$ with $a=1,2$ will be calculated below; here we transform the Keldysh matrix form of
term $S^{(0)}$ into more convenient representation via classical (cl) and quantum (q) components:
\begin{eqnarray}
\label{S0}
S^{(0)}_{\tilde{\Phi}}=\frac{ig}{ \varepsilon_0}\int \frac{d\Omega}{2\pi} \hat{\tilde{\Phi}}_{-\Omega}^T
\hat{L}(\Omega)  \hat{\tilde{\Phi}}_{\Omega} \quad \hat{\tilde{\Phi}}_{\Omega} =\left(\begin{smallmatrix}
\tilde{\Phi}^{cl}_{\Omega} \\ \tilde{\Phi}^{q}_{\Omega}
\end{smallmatrix}\right) \quad \tilde{\Phi}^{\pm}=\tilde{\Phi}^{cl}\pm\tilde{\Phi}^{q}\quad \\ \nonumber
L(\Omega)=-2i\left(\begin{smallmatrix}
0 &&  K(-\Omega) \sin(\pi d)\\ K(\Omega) \sin(\pi d) && i\cos(\pi d) \left( K(-\Omega)+K(\Omega)\right)
\end{smallmatrix}\right)\nonumber \\
\end{eqnarray}
Eq.(\ref{S0}) provides us with convenient form of $S^{(0)}$ to be used below. In the following two subsections we 
calculate  $S^{(1)}$ and $S^{(2)}$.

This expression also helps us to calculate contribution to the heat capacity from the perturbation on the mean-field level. The expression for the heat capacity takes the form:
\begin{eqnarray}
C=C_{SYK}+\frac{\partial}{\partial T} \left(\frac{1}{2}\frac{\delta S^{(0)}_{\tilde{\Phi}}  }{\delta \Phi^{q}(t)}|_{\Phi=0}\right)=(2\pi)^2 \alpha_S N\frac{T}{J}+\frac{\partial}{\partial_T}\frac{2\pi}{\beta}\frac{ig}{ 2\varepsilon_0} 
L_{q,cl}(0)  
\end{eqnarray}
Taking the limit $d\rightarrow \frac{1}{2}$ we meet divergence which  is present since $g^2(x)\propto |x|^{-1}$ at $x\rightarrow 0$. We need to cut-off this integral at the scale $x\propto \frac{T}{J}$. As a result we have:
\begin{eqnarray}
\label{C01}
C= (2\pi)^2 \alpha_S N\frac{T}{J}-\frac{\partial}{\partial_T}\frac{2\pi}{\beta}\frac{g}{ \varepsilon_0} \ln(\beta J)= (2\pi)^2 \alpha_S N \frac{T}{J}+\sqrt{b}\frac{ N\Gamma^2}{2 T J} 
\end{eqnarray}


\subsection{Calculation of $S^{(1)}$ }
\label{App_2_3}
To calculate $S^{(1)}$ we represent it in the form
\begin{eqnarray}
S^{(1)}_{\tilde{\Phi}}=\frac{ig}{2 \varepsilon_0}\sum\limits_{s_1,s_0}\int dx_1 dx_0 \int \frac{d\Omega_1}{2\pi} e^{-i \Omega_1 (x_1-x_0) } \hat{L}_{s_1,s_0}(\Omega_1)\left(du^{s_1\prime}(x_1)-i \Omega_1 u^{s_1}(x_1)\right)  \tilde{\Phi}^{s_1}(x_1)\tilde{\Phi}^{s_0}(x_0)+\nonumber \\\frac{ig}{2 \varepsilon_0}\sum\limits_{s_1,s_0}\int dx_1 dx_0 \int \frac{d\Omega_1}{2\pi} e^{-i \Omega_1 (x_1-x_0) } \hat{L}_{s_1,s_0}(\Omega_1) \left(du^{s_0\prime}(x_0)+i \Omega_1 u^{s_0}(x_0)\right)  \tilde{\Phi}^{s_1}(x_1)\tilde{\Phi}^{s_0}(x_0) \nonumber \\
\end{eqnarray}
Now we need to substitute Fourier representation of $u(x)$, and also take into account the symmetry w.r.t.  swap  of times $x_1$ and $x_2$; in the result,
\begin{eqnarray}
S^{(1)}_{\tilde{\Phi}}=\frac{g}{ \varepsilon_0}\sum\limits_{s_1,s_0}\int dx_1 dx_0 \int \frac{d\Omega_1d\Omega}{(2\pi)^2} e^{-i \Omega_1 (x_1-x_0) } \hat{L}_{s_1,s_0}(\Omega_1)\left(\Omega d+ \Omega_1\right)e^{-i\Omega x_1} u_{\Omega}^{s_1} \tilde{\Phi}^{s_1}(x_1)\tilde{\Phi}^{s_0}(x_0). \nonumber \\
\end{eqnarray}

To make expression more symmetric we perform shift  $\Omega_1\rightarrow \Omega_1-\frac{\Omega}{2}$, and then integrate
over $x_i$:
\begin{eqnarray}
\label{S1}
S^{(1)}_{\tilde{\Phi}}=
\frac{g}{ \varepsilon_0}\sum\limits_{s_1,s_0} \int \frac{d\Omega_1d\Omega}{(2\pi)^2}  \hat{L}_{s_1,s_0}\left(\Omega_1-\frac{\Omega}{2}\right)\left(\Omega  (d-\frac{1}{2})+ \Omega_1\right) u_{\Omega}^{s_1} \tilde{\Phi}^{s_1}_{-\Omega_1-\frac{\Omega}{2}}\tilde{\Phi}^{s_0}_{\Omega_1-\frac{\Omega}{2}} 
\end{eqnarray}
Finally, we transform (\ref{S1}) into the $(cl,q)$ representation:
\begin{eqnarray}
\label{S11}
S^{(1)}= 
\frac{2g}{ \varepsilon_0}\int \frac{d\Omega_1d\Omega}{(2\pi)^2}  \frac{1}{\sqrt{2}}\left( \begin{smallmatrix}
u^{cl}_{\Omega} \\ u^{q}_{\Omega}
\end{smallmatrix}\right)^T \left(\begin{smallmatrix}
\hat{\tilde{\Phi}}_{-\Omega_1-\frac{\Omega}{2}}^T L_{I,1}(\Omega,\Omega_1) \hat{\tilde{\Phi}}_{\Omega_1-\frac{\Omega}{2}} \\ \hat{\tilde{\Phi}}_{-\Omega_1-\frac{\Omega}{2}}^T\tau_X L_{I,1}(\Omega,\Omega_1)\hat{\tilde{\Phi}}_{\Omega_1-\frac{\Omega}{2}}
\end{smallmatrix}\right)\quad L_{I,1}(\Omega,\Omega_1)=\left(\Omega  (d-\frac{1}{2})+ \Omega_1\right)\hat{L}\left(\Omega_1-\frac{\Omega}{2}\right)\nonumber\\
\end{eqnarray}
The last formula concludes our derivation. Here $\tau_X$ is a Pauli matrix $X$ in the space $cl,q$.

\subsection{Calculation of $S^{(2)}$ }
\label{App_2_3:Calculation of $S^{(2)}$}
The second-order terms consist of two groups which will be calculated in sequence.

\subsubsection{Calculation of $S^{(2,1)}$}
\label{App_2_3_1}
After the use of symmetry to the time swap, the expression for $S^{(2,1)}$ reads:
\begin{eqnarray}
S^{(2,1)}_{\tilde{\Phi}}=\frac{ig}{ \varepsilon_0}\sum\limits_{s_1,s_0}\int dx_1 dx_0 \int \frac{d\Omega}{2\pi} e^{-i \Omega (x_1-x_0) } \hat{L}_{s_1,s_0}(\Omega)   \nonumber \\ \left( \frac{d(d-1)}{2}\left([u^{s_1\prime}(x_1)]^2\right) -\frac{\Omega^2}{2}\left( [u^{s_1}(x_1)]^2\right)-i \Omega d (u^{s_1}(x_1)u^{s_1\prime}(x_1)) \right)\tilde{\Phi}^{s_1}(x_1)\tilde{\Phi}^{s_0}(x_0)\quad 
\end{eqnarray}
We proceed similar to the previous case. After Fourier transformation  of $u(x)$ we perform the frequency shift and do
the integrals over $x_i$; as a result we obtain:
\begin{eqnarray}
S^{(2,1)}_{\tilde{\Phi}}=-\frac{ig}{ \varepsilon_0}\sum\limits_{s_1,s_0} \int \frac{d\Omega_1d\Omega_2d\Omega_2}{(2\pi)^3}     \left( \frac{d(d-1)}{2}\Omega_0 \Omega_1 +\frac{1}{2}\left(\Omega_2-\frac{\Omega_1+\Omega_0}{2}\right)^2+ d\left(\Omega_2-\frac{\Omega_1+\Omega_0}{2}\right)  \Omega_1 \right)\nonumber \\ \hat{L}_{s_1,s_0}\left(\Omega_2-\frac{\Omega_1+\Omega_0}{2}\right) u^{s_1}_{\Omega_0}u^{s_1}_{\Omega_1 }\tilde{\Phi}^{s_1}_{-\Omega_2-\frac{\Omega_1+\Omega_0}{2}}\tilde{\Phi}^{s_0}_{\Omega_2-\frac{\Omega_1+\Omega_0}{2}}
\end{eqnarray}
Finally we present results in the matrix form: 
\begin{eqnarray}
S^{(2,1)}_{\tilde{\Phi}}=
-\frac{ig}{ \varepsilon_0} \int \frac{d\Omega_1d\Omega_2d\Omega_2}{(2\pi)^3}     \left( \frac{d(d-1)}{2}\Omega_0 \Omega_1 +\frac{1}{2}\left(\Omega_2-\frac{\Omega_1+\Omega_0}{2}\right)^2+ d\left(\Omega_2-\frac{\Omega_1+\Omega_0}{2}\right)  \Omega_1 \right) \nonumber \\
\hat{u}_{\Omega_0}^T\left( \begin{smallmatrix}
\hat{\tilde{\Phi}}_{-\Omega_2-\frac{\Omega_1+\Omega_0}{2}}^T\hat{L}\left(\Omega_2-\frac{\Omega_1+\Omega_0}{2}\right) \hat{\tilde{\Phi}}_{\Omega_2-\frac{\Omega_1+\Omega_0}{2}} && \hat{\tilde{\Phi}}_{-\Omega_2-\frac{\Omega_1+\Omega_0}{2}}^T\tau_X\hat{L}\left(\Omega_2-\frac{\Omega_1+\Omega_0}{2}\right) \hat{\tilde{\Phi}}_{\Omega_2-\frac{\Omega_1+\Omega_0}{2}} \\ \hat{\tilde{\Phi}}_{-\Omega_2-\frac{\Omega_1+\Omega_0}{2}}^T\tau_X\hat{L}\left(\Omega_2-\frac{\Omega_1+\Omega_0}{2}\right) \hat{\tilde{\Phi}}_{\Omega_2-\frac{\Omega_1+\Omega_0}{2}} && \hat{\tilde{\Phi}}_{-\Omega_2-\frac{\Omega_1+\Omega_0}{2}}^T\hat{L}\left(\Omega_2-\frac{\Omega_1+\Omega_0}{2}\right) \hat{\tilde{\Phi}}_{\Omega_2-\frac{\Omega_1+\Omega_0}{2}}
\end{smallmatrix} \right)\hat{u}_{\Omega_1}^T  \nonumber \\
\end{eqnarray}
It concludes our calculation of $S^{(2,1)}$.

\subsubsection{Calculation of $S^{(2,2)}$}
\label{App_2_3_2}
This term can be written in the  Fourier domain after integration over $x_i$:
\begin{eqnarray}
S^{(2,2)}_{\tilde{\Phi}}=-\frac{ig}{2 \varepsilon_0}\sum\limits_{s_1,s_0}\int \frac{d\Omega_2d\Omega_0d\Omega_1}{(2\pi)^3}\left(d^2\Omega_0\Omega_1 -\left(\Omega_2-\frac{\Omega_1-\Omega_0}{2}\right)^2+ d\left(\Omega_2-\frac{\Omega_1-\Omega_0}{2}\right) (\Omega_0-\Omega_1) \right) \nonumber \\
u^{s_1}_{\Omega_1}u^{s_0}_{\Omega_0}  \hat{L}_{s_1,s_0}\left(\Omega_2-\frac{\Omega_1-\Omega_0}{2}\right) \tilde{\Phi}^{s_1}_{-\Omega_2-\frac{\Omega_1+\Omega_0}{2}}\tilde{\Phi}^{s_0}_{\Omega_2-\frac{\Omega_1+\Omega_0}{2}}
\end{eqnarray}
Writing this expression in $(cl,q)$ notations, we find:
\begin{eqnarray}
S^{(2,2)}_{\tilde{\Phi}}=-\frac{ig}{2 \varepsilon_0}\int \frac{d\Omega_2d\Omega_0d\Omega_1}{(2\pi)^3}\left(d^2\Omega_0\Omega_1 -\left(\Omega_2-\frac{\Omega_1-\Omega_0}{2}\right)^2+ d\left(\Omega_2-\frac{\Omega_1-\Omega_0}{2}\right) (\Omega_0-\Omega_1) \right) \nonumber \\
\hat{u}_{\Omega_1}^T \left(\begin{smallmatrix}
\hat{\tilde{\Phi}}_{-\Omega_2-\frac{\Omega_1+\Omega_0}{2}}^T\hat{L}\left(\Omega_2-\frac{\Omega_1-\Omega_0}{2}\right)\hat{\tilde{\Phi}}_{\Omega_2-\frac{\Omega_1+\Omega_0}{2}} && \hat{\tilde{\Phi}}_{-\Omega_2-\frac{\Omega_1+\Omega_0}{2}}^T\hat{L}\left(\Omega_2-\frac{\Omega_1-\Omega_0}{2}\right)\tau_X\hat{\tilde{\Phi}}_{\Omega_2-\frac{\Omega_1+\Omega_0}{2}} \\ \hat{\tilde{\Phi}}_{-\Omega_2-\frac{\Omega_1+\Omega_0}{2}}^T\tau_X\hat{L}\left(\Omega_2-\frac{\Omega_1-\Omega_0}{2}\right)\hat{\tilde{\Phi}}_{\Omega_2-\frac{\Omega_1+\Omega_0}{2}} && \hat{\tilde{\Phi}}_{-\Omega_2-\frac{\Omega_1+\Omega_0}{2}}^T\tau_X\hat{L}\left(\Omega_2-\frac{\Omega_1-\Omega_0}{2}\right)\tau_X\hat{\tilde{\Phi}}_{\Omega_2-\frac{\Omega_1+\Omega_0}{2}}
\end{smallmatrix} \right) \hat{u}_{\Omega_0} 
\end{eqnarray}
\subsubsection{Summation of two parts of $S^{(2)}$}
\label{App_2_3_3}
Finally, we calculate $S^{(2)} = S^{(2,1)} + S^{(2,2)}$. Both expressions for $S^{(2,1)}$ and $S^{(2,2)}$ share
common matrix structure, which contains two $2\times 2$ Keldysh spaces,each one for $u(x)$ and $\Phi(x)$ variables.
Their sum can represented in the following way in terms of 4 new matrices of rank 2 each:
\begin{eqnarray}
&\left(\begin{smallmatrix}
\hat{L}^Q_{II}(\Omega_0,\Omega_1,\Omega_2) && \hat{L}^A_{II}(\Omega_0,\Omega_1,\Omega_2) \\ \hat{L}^R_{II}(\Omega_0,\Omega_1,\Omega_2) &&
\hat{L}^K_{II}(\Omega_0,\Omega_1,\Omega_2)
\end{smallmatrix}\right)\equiv  \nonumber \\ &    \left( d(d-1)\Omega_0 \Omega_1 +\left(\Omega_2-\frac{\Omega_1+\Omega_0}{2}\right)^2+ 2d\left(\Omega_2-\frac{\Omega_1+\Omega_0}{2}\right)  \Omega_0 \right) \left( \begin{smallmatrix}
\hat{L}\left(\Omega_2-\frac{\Omega_1+\Omega_0}{2}\right) && \tau_X\hat{L}\left(\Omega_2-\frac{\Omega_1+\Omega_0}{2}\right)  \\ \tau_X\hat{L}\left(\Omega_2-\frac{\Omega_1+\Omega_0}{2}\right) && \hat{L}\left(\Omega_2-\frac{\Omega_1+\Omega_0}{2}\right) 
\end{smallmatrix} \right)+ \nonumber \\
&\left(d^2\Omega_0\Omega_1 -\left(\Omega_2-\frac{\Omega_1-\Omega_0}{2}\right)^2+ d\left(\Omega_2-\frac{\Omega_1-\Omega_0}{2}\right) (\Omega_0-\Omega_1) \right)\left(\begin{smallmatrix}
\hat{L}\left(\Omega_2-\frac{\Omega_1-\Omega_0}{2}\right) &&  \hat{L}\left(\Omega_2-\frac{\Omega_1-\Omega_0}{2}\right)\tau_X \\ \tau_X\hat{L}\left(\Omega_2-\frac{\Omega_1-\Omega_0}{2}\right) && \tau_X\hat{L}\left(\Omega_2-\frac{\Omega_1-\Omega_0}{2}\right)\tau_X
\end{smallmatrix} \right)
\label{eq: L_II_full} \nonumber\\
\end{eqnarray}
As a result we can write $S^{(2)}$ as 
\begin{eqnarray}
S^{(2)}_{\tilde{\Phi}}=-\frac{ig}{2 \varepsilon_0}\int \frac{d\Omega_2d\Omega_0d\Omega_1}{(2\pi)^3}  \hat{u}_{\Omega_1}^T \left(\begin{smallmatrix}
\hat{\tilde{\Phi}}_{-\Omega_2-\frac{\Omega_1+\Omega_0}{2}}^T\hat{L}^Q_{II}(\Omega_0,\Omega_1,\Omega_2)\hat{\tilde{\Phi}}_{\Omega_2-\frac{\Omega_1+\Omega_0}{2}} && \hat{\tilde{\Phi}}_{-\Omega_2-\frac{\Omega_1+\Omega_0}{2}}^T\hat{L}^A_{II}(\Omega_0,\Omega_1,\Omega_2)\hat{\tilde{\Phi}}_{\Omega_2-\frac{\Omega_1+\Omega_0}{2}} \\ \hat{\tilde{\Phi}}_{-\Omega_2-\frac{\Omega_1+\Omega_0}{2}}^T\hat{L}^R_{II}(\Omega_0,\Omega_1,\Omega_2)\hat{\tilde{\Phi}}_{\Omega_2-\frac{\Omega_1+\Omega_0}{2}} && \hat{\tilde{\Phi}}_{-\Omega_2-\frac{\Omega_1+\Omega_0}{2}}^T\hat{L}^K_{II}(\Omega_0,\Omega_1,\Omega_2)\hat{\tilde{\Phi}}_{\Omega_2-\frac{\Omega_1+\Omega_0}{2}}
\end{smallmatrix} \right) \hat{u}_{\Omega_0} 
\label{eq:S^{(2)}}
\end{eqnarray}
This action leads both the modification of the effective action of soft modes, and to additional contributions 
to the susceptibility.

\section{Quadratic action of soft modes.}
\label{App_3}
In this section we will show that the combined SYK$_4$ SYK$_2$ model with $\Gamma\gg T$ demonstrates properties which
are very different from the pure SYK model. To find the  correction to the action of soft mode due to $\Gamma_{ij}$ terms
in the Hamiltonian, we  set   $\hat{\tilde{\Phi}}_{\Omega}=\Phi^{cl}_{\Omega} \left(\begin{smallmatrix}
1 \\0
\end{smallmatrix}\right)$. Using this substitution in (\ref{eq:S^{(2)}}) we can write:
\begin{eqnarray}
S^{(2)}_{\tilde{\Phi}}=-\frac{ig}{2 \varepsilon_0}\int \frac{d\Omega_2d\Omega_0d\Omega_1}{(2\pi)^3}\tilde{\Phi}^{cl}_{-\Omega_2-\frac{\Omega_1+\Omega_0}{2}}   \tilde{\Phi}^{cl}_{\Omega_2-\frac{\Omega_1+\Omega_0}{2}}\hat{u}_{\Omega_1}^T \hat{L}_{II}(\Omega_0,\Omega_1,\Omega_2)\hat{u}_{\Omega_0}\nonumber \\ \hat{L}_{II}(\Omega_0,\Omega_1,\Omega_2)= \left(\begin{smallmatrix}
\left[\hat{L}^Q_{II}(\Omega_0,\Omega_1,\Omega_2)\right]_{cl,cl} && \left[\hat{L}^A_{II}(\Omega_0,\Omega_1,\Omega_2)\right]_{cl,cl} \\ \left[\hat{L}^R_{II}(\Omega_0,\Omega_1,\Omega_2)\right]_{cl,cl} &&
\left[\hat{L}^K_{II}(\Omega_0,\Omega_1,\Omega_2)\right]_{cl,cl}
\end{smallmatrix}\right)
\end{eqnarray}
and  the consequencies which come due to this additional term in the effective action.

\subsection{Action for the problem without pumping}
\label{App_3_1}
In the absense of pumping   $\tilde{\Phi}^{cl}_{\Omega}=2\pi \delta(\Omega)$ and we get  the following simple expression
for the additional action:
\begin{eqnarray}
\delta S_{soft}=-\frac{ig}{2 \varepsilon_0}\int \frac{d\Omega}{2\pi}\hat{u}_{-\Omega}^T \hat{L}_{II}(\Omega,-\Omega,0)\hat{u}_{\Omega}
\end{eqnarray}
Here it is  time  to wire $\hat{L}_{II}(\Omega,-\Omega,0)$ explicitly:
\begin{eqnarray}
\boxed{
	\hat{L}_{II}(\Omega,-\Omega,0) = -\frac{i \Omega^2}{2}\left(\begin{smallmatrix}
	0 && \psi(\Omega) \\ \psi(-\Omega) && -i\pi
	\end{smallmatrix} \right)\quad \psi(\Omega)=\Psi\left(\frac{1}{2}+i\Omega\right)-\Psi\left(-\frac{1}{2}\right)\quad \Psi(z)=\partial_z \ln \Gamma(z)}
\end{eqnarray}
The full action  quadratic action of the model has form:
\begin{eqnarray}
&\delta S_{soft}=\frac{1}{2}\int \frac{d\Omega}{2\pi}\hat{u}_{-\Omega}^T \left(\begin{smallmatrix}
0 && \left[\mathcal{G}^A(\Omega)\right]^{-1} \\ \left[\mathcal{G}^R(\Omega)\right]^{-1} && -\mathcal{G}^K(\Omega) \left[\mathcal{G}^A(\Omega)\right]^{-1}\left[\mathcal{G}^R(\Omega)\right]^{-1}
\end{smallmatrix} \right) \hat{u}_{\Omega}\nonumber \\ &\left[\mathcal{G}^{R(A)}(\Omega)\right]^{-1}=\Omega^2 \left(\varepsilon_0 \left(\Omega^2+1\right)-\frac{g}{2\varepsilon_0}\psi(\pm\Omega)\right)\quad -\mathcal{G}^K(\Omega) \left[\mathcal{G}^A(\Omega)\right]^{-1}\left[\mathcal{G}^R(\Omega)\right]^{-1}=i\frac{g}{2\varepsilon_0}\pi \nonumber\\
\end{eqnarray}
One can makes two observations: first, the distribution function determined by the relation $\mathcal{G}^K(\Omega)\equiv F(\Omega)\left(\mathcal{G}^R(\Omega)-\mathcal{G}^A(\Omega)\right)$  reaches its equilibrium value: $F(\Omega)=\coth(\pi \Omega)$. Secondly, for  $g\gg \varepsilon^2_0$, we  observe the "resonant" behavior with the resonance frequency $\Omega_R\gg1$. The position of the resonance and the behavior of the Green  function in its vicinity are determined by the relations:
\begin{eqnarray}
\varepsilon_0 \left(\Omega_R^2+1\right)=\frac{g}{2\varepsilon_0}\Re\psi(\Omega_R)\quad \mathcal{G}^{R(A)}(\Omega)\approx \frac{1}{2\Omega_R^3 \varepsilon_0}\frac{sgn(\Omega)}{\delta\Omega\pm i\Omega_W}\nonumber \\ \Omega= \pm\Omega_R+\delta \Omega\quad \frac{\Omega_W}{\Omega_R}= -\frac{1}{2}\frac{\Im\psi(\Omega_R) }{\Re\psi(\Omega_R)}\approx\frac{\pi}{4\ln(\Omega_R)} \nonumber \\
\end{eqnarray}

\subsection{Action for the problem with pumping}
\label{App_3_2}
In the problem with pumping we have $\tilde{\Phi}_{\Omega}=2\pi\left[\delta(\Omega)+\frac{A}{2}\left(\delta(\Omega-\Omega_P)+
\delta(\Omega + \Omega_P)\right)\right]$. 
In the previous subsection we have described the term without $A$. The term linear in $A$ has the form:
\begin{eqnarray}
\delta^{(1)} S_{soft}=-\frac{ig}{2 \varepsilon_0}\frac{A}{2}\int \frac{d\Omega}{(2\pi)}  \hat{u}_{-\Omega-\Omega_P}^T L_{P,I,1}(\Omega,\Omega_P) \hat{u}_{\Omega}+ \left[\Omega_P\rightarrow-\Omega_P \right] \nonumber \\
L_{P,I,1}(\Omega,\Omega_P)=\left[\hat{L}_{II}(\Omega,-\Omega-\Omega_P,\frac{\Omega_P}{2})+\hat{L}_{II}(\Omega,-\Omega-\Omega_P,-\frac{\Omega_P}{2})\right] \nonumber \\
\end{eqnarray}
We see that slow and fast modes of $u(x)$ fluctuations are coupled due to the pumping term.
We  integrate now over fast motions to find action for slow mode alone:
\begin{eqnarray}
\label{dS21}
\delta^{(2,1)} S_{soft}=\frac{i}{2}\langle  \left[\delta^{(1)} S_{soft}\right]^2 \rangle_{fast}= \nonumber \\ \frac{1}{2}\left[\frac{g}{2 \varepsilon_0}\frac{A}{2}\right]^2 \int\limits_{\Omega \ll \Omega_W} \frac{d\Omega}{(2\pi)} \hat{u}_{-\Omega}^T  L_{P,I}(-\Omega,-\Omega_P)^T \mathcal{G}_{\Omega_P} L_{P,I}(\Omega,\Omega_P) \hat{u}_{\Omega}+
\left[\Omega_P\rightarrow-\Omega_P \right] \nonumber \\
\end{eqnarray}
where
\begin{equation}
L_{P,I}(\Omega,\Omega_P)=\frac{1}{2}\left( L_{P,I,1}(\Omega,\Omega_P)+ L_{P,I,1}(-\Omega-\Omega_P,\Omega_P)^T \right)
\end{equation}
In the main order over $\Omega/\Omega_W \ll 1$ the action in Eq.(\ref{dS21}) has the form:
\begin{eqnarray}
\delta^{(2,1)} S_{soft}\approx  \pi \Omega_R \frac{ig}{2 \varepsilon_0} \left(\frac{A}{2}\right)^2 \int\limits_{\Omega \ll \Omega_W} \frac{d\Omega}{(2\pi)} \hat{u}_{-\Omega}^T  \left(\begin{smallmatrix}
0 && \Omega\\ -\Omega && 0
\end{smallmatrix}\right) \hat{u}_{\Omega} 
\end{eqnarray}

The  term quadratic in $A$ leads to the following correction to the action:
\begin{eqnarray}
\delta^{(2,2)} S_{soft}\approx=-\frac{ig}{2 \varepsilon_0}\left(\frac{A}{2}\right)^2\int \frac{d\Omega}{(2\pi)} \hat{u}_{-\Omega}^T \left[\hat{L}_{II}(\Omega,-\Omega,-\Omega_P)+ \left(\Omega_P\rightarrow  -\Omega_P\right) \right]\hat{u}_{\Omega} \approx \nonumber \\\pi \Omega_P\frac{ig}{2 \varepsilon_0}\left(\frac{A}{2}\right)^2\int \frac{d\Omega}{(2\pi)} \hat{u}_{-\Omega}^T \left(\begin{smallmatrix}
0 && -2\Omega \\ 2\Omega && 4 \Omega_P
\end{smallmatrix}\right)\hat{u}_{\Omega}
\end{eqnarray}
The combination of  both contributions gives:
\begin{eqnarray}
\delta_A S_{soft}=\pi \Omega_P\frac{ig}{2 \varepsilon_0}\left(\frac{A}{2}\right)^2\int \frac{d\Omega}{(2\pi)} \hat{u}_{-\Omega}^T \left(\begin{smallmatrix}
0 && -\Omega \\ \Omega && 4 \Omega_P
\end{smallmatrix}\right)\hat{u}_{\Omega}
\end{eqnarray}

\section{Calculation of susceptibility}
\label{App_4}
Relative smallness of $u(x)$ fluctuations allows (as explained in the main text) to represent
the susceptibility has the following form:
\begin{eqnarray}
\chi(\Omega)=\frac{1}{2}\frac{\delta^2 }{\delta f^{q}_{\Omega}\delta f^{cl}_{\Omega}}\left(S^{(0)}_{\tilde{\Phi}}+\frac{i}{2}\langle [S^{(1)}]^2  \rangle\right)
\end{eqnarray}
The calculation of $S^{(0)}$  was performed above. It is time to calculate the second term.

\subsection{Calculation of $\frac{i}{2}\langle [S^{(1)}]^2  \rangle$}
\label{App_4_1}
Using the expression for $S^{(1)}$ from Eq.(\ref{S11}) and doing the averaging, we obtain
\begin{eqnarray}
&\frac{i}{2}\langle [S^{(1)}]^2  \rangle= i \left(\frac{g}{ \varepsilon_0}\right)^2\langle 
\int \frac{d\Omega_1d\Omega}{(2\pi)^2}\frac{d\Omega_2d\Omega^\prime}{(2\pi)^2} \left(\begin{smallmatrix}
\hat{\tilde{\Phi}}_{-\Omega_1-\frac{\Omega}{2}}^T L_{I,1}(\Omega,\Omega_1) \hat{\tilde{\Phi}}_{\Omega_1-\frac{\Omega}{2}} \\ \hat{\tilde{\Phi}}_{-\Omega_1-\frac{\Omega}{2}}^T\tau_X L_{I,1}(\Omega,\Omega_1)\hat{\tilde{\Phi}}_{\Omega_1-\frac{\Omega}{2}}
\end{smallmatrix}\right)^T \left( \begin{smallmatrix}
u^{cl}_{\Omega} \\ u^{q}_{\Omega}
\end{smallmatrix}\right)
\left( \begin{smallmatrix}
u^{cl}_{\Omega^\prime} \\ u^{q}_{\Omega^\prime}
\end{smallmatrix}\right)^T \left(\begin{smallmatrix}
\hat{\tilde{\Phi}}_{-\Omega_2-\frac{\Omega^\prime}{2}}^T L_{I,1}(\Omega^\prime,\Omega_2) \hat{\tilde{\Phi}}_{\Omega_2-\frac{\Omega^\prime}{2}} \\ \hat{\tilde{\Phi}}_{-\Omega_2-\frac{\Omega^\prime}{2}}^T\tau_X L_{I,1}(\Omega^\prime,\Omega_2)\hat{\tilde{\Phi}}_{\Omega_2-\frac{\Omega^\prime}{2}}
\end{smallmatrix}\right) \rangle  = \nonumber \\
& - \left(\frac{g}{ \varepsilon_0}\right)^2 
\int \frac{d\Omega_1d\Omega d\Omega_2}{(2\pi)^3} \left(\begin{smallmatrix}
\hat{\tilde{\Phi}}_{-\Omega_1-\frac{\Omega}{2}}^T L_{I,1}(\Omega,\Omega_1) \hat{\tilde{\Phi}}_{\Omega_1-\frac{\Omega}{2}} \\ \hat{\tilde{\Phi}}_{-\Omega_1-\frac{\Omega}{2}}^T\tau_X L_{I,1}(\Omega,\Omega_1)\hat{\tilde{\Phi}}_{\Omega_1-\frac{\Omega}{2}}
\end{smallmatrix}\right)^T  \hat{\mathcal{G}}_{\Omega} \left(\begin{smallmatrix}
\hat{\tilde{\Phi}}_{-\Omega_2+\frac{\Omega}{2}}^T L_{I,1}(-\Omega,\Omega_2) \hat{\tilde{\Phi}}_{\Omega_2+\frac{\Omega}{2}} \\ \hat{\tilde{\Phi}}_{-\Omega_2+\frac{\Omega}{2}}^T\tau_X L_{I,1}(-\Omega,\Omega_2)\hat{\tilde{\Phi}}_{\Omega_2+\frac{\Omega}{2}}
\end{smallmatrix}\right) 
\end{eqnarray}
Here we used the relation $-i\langle \left( \begin{smallmatrix}
u^{cl}_{\Omega} \\ u^{q}_{\Omega}
\end{smallmatrix}\right)
\left( \begin{smallmatrix}
u^{cl}_{\Omega^\prime} \\ u^{q}_{\Omega^\prime}
\end{smallmatrix}\right)^T \rangle =2\pi \delta(\Omega+\Omega^\prime)\hat{\mathcal{G}}_{\Omega}$
where 
$ \hat{\mathcal{G}}_{\Omega}=\left(\begin{smallmatrix}
\mathcal{G}^K_{\Omega} && \mathcal{G}^R_{\Omega} \\ \mathcal{G}^A_{\Omega} && 0
\end{smallmatrix}\right)$. 
To calculate susceptibility we need  to introduce a probe field by replacing $\tilde{\Phi}$ by $ \tilde{\Phi}+f$ after that we should find a second order term in $f$; after that  we can use $\tilde{\Phi}$ containing classical component only. We denote the
needed quadratic  term as $\left\{\frac{i}{2}\langle [S^{(1)}]^2  \rangle\right\}^{(2)}_f$  and calculate it now.

\subsection{Calculation of  $\left\{\frac{i}{2}\langle [S^{(1)}]^2  \rangle\right\}^{(2)}_f$}
\label{App_4_2}
According to the definition:
\begin{eqnarray}
&\left\{\frac{i}{2}\langle [S^{(1)}]^2  \rangle\right\}^{(2)}_f=  - \left(\frac{g}{ \varepsilon_0}\right)^2 
\int \frac{d\Omega_1d\Omega d\Omega_2}{(2\pi)^3} \left(\begin{smallmatrix}
\hat{\tilde{\Phi}}_{-\Omega_1-\frac{\Omega}{2}}^T L_{I,2,1}(\Omega,\Omega_1)\hat{f}_{\Omega_1-\frac{\Omega}{2}}\\ \hat{\tilde{\Phi}}_{-\Omega_1-\frac{\Omega}{2}}^T L_{I,2,2}(\Omega,\Omega_1)\hat{f}_{\Omega_1-\frac{\Omega}{2}}
\end{smallmatrix}\right)^T  \hat{\mathcal{G}}_{\Omega} \left(\begin{smallmatrix}
\hat{\tilde{\Phi}}_{-\Omega_2+\frac{\Omega}{2}}^T L_{I,2,1}(-\Omega,\Omega_2)\hat{f}_{\Omega_2+\frac{\Omega}{2}} \\ \hat{\tilde{\Phi}}_{-\Omega_2+\frac{\Omega}{2}}^TL_{I,2,2}(-\Omega,\Omega_2)\hat{f}_{\Omega_2+\frac{\Omega}{2}}
\end{smallmatrix}\right) + \nonumber \\
&- 2\left(\frac{g}{ \varepsilon_0}\right)^2 
\int \frac{d\Omega_1d\Omega d\Omega_2}{(2\pi)^3}  \left(\begin{smallmatrix}
\hat{f}_{-\Omega_1-\frac{\Omega}{2}}^T L_{I,2,1}(\Omega,\Omega_1)\hat{f}_{\Omega_1-\frac{\Omega}{2}}\\ \hat{f}_{-\Omega_1-\frac{\Omega}{2}}^T L_{I,2,2}(\Omega,\Omega_1)\hat{f}_{\Omega_1-\frac{\Omega}{2}}
\end{smallmatrix}\right)^T  \hat{\mathcal{G}}_{\Omega} \left(\begin{smallmatrix}
\hat{\tilde{\Phi}}_{-\Omega_2+\frac{\Omega}{2}}^T L_{I,1}(-\Omega,\Omega_2) \hat{\tilde{\Phi}}_{\Omega_2+\frac{\Omega}{2}} \\ \hat{\tilde{\Phi}}_{-\Omega_2+\frac{\Omega}{2}}^T\tau_X L_{I,1}(-\Omega,\Omega_2)\hat{\tilde{\Phi}}_{\Omega_2+\frac{\Omega}{2}}
\end{smallmatrix}\right) 
\end{eqnarray}
Here $L_{I,2,1}(\Omega,\Omega_1)=L_{I,1}(\Omega,\Omega_1)+L_{I,1}(\Omega,-\Omega_1)^T$ and $L_{I,2,2}(\Omega,\Omega_1)=\tau_X L_{I,1}(\Omega,\Omega_1)+ L_{I,1}(\Omega,-\Omega_1)^T \tau_X$. Finally, we need to write this term in more convenient form (also we did not use the classical structure of the field $\tilde{\Phi}$)
\begin{eqnarray}
&\left\{\frac{i}{2}\langle [S^{(1)}]^2  \rangle\right\}^{(2)}_f=  - \left(\frac{g}{ \varepsilon_0}\right)^2 
\int \frac{d\Omega_1d\Omega d\Omega_2}{(2\pi)^3}\tilde{\Phi}^{cl}_{-\Omega_1-\frac{\Omega}{2}} \tilde{\Phi}^{cl}_{-\Omega_2+\frac{\Omega}{2}}\hat{f}_{\Omega_1-\frac{\Omega}{2}}^{T}\hat{\Pi}_{I,1}(\Omega,\Omega_1,\Omega_2) \hat{f}_{\Omega_2+\frac{\Omega}{2}}+\nonumber \\ 
&- 2\left(\frac{g}{ \varepsilon_0}\right)^2 
\int \frac{d\Omega_1d\Omega d\Omega_2}{(2\pi)^3}  \tilde{\Phi}^{cl}_{\Omega_2+\frac{\Omega}{2}} \tilde{\Phi}^{cl}_{-\Omega_2+\frac{\Omega}{2}}\hat{f}_{-\Omega_1-\frac{\Omega}{2}}^T  \hat{\Pi}_{I,2}(\Omega,\Omega_1,\Omega_2)\hat{f}_{\Omega_1-\frac{\Omega}{2}}
\nonumber \\
&\left[\hat{\Pi}_{I,1}(\Omega,\Omega_1,\Omega_2)\right]_{s_1,s_0}=\left(\begin{smallmatrix}
\left[L_{I,2,1}(\Omega,\Omega_1)\right]_{cl,s_1}\\ \left[ L_{I,2,2}(\Omega,\Omega_1)\right]_{cl,s_1}
\end{smallmatrix}\right)^T  \hat{\mathcal{G}}_{\Omega} \left(\begin{smallmatrix}
\left[L_{I,2,1}(-\Omega,\Omega_2)\right]_{cl,s_0} \\ \left[L_{I,2,2}(-\Omega,\Omega_2)\right]_{cl,s_0}
\end{smallmatrix}\right) \nonumber \\
& \hat{\Pi}_{I,2}(\Omega,\Omega_1,\Omega_2) =\left(\begin{smallmatrix}
L_{I,2,1}(\Omega,\Omega_1)\\  L_{I,2,2}(\Omega,\Omega_1)
\end{smallmatrix}\right)^T  \hat{\mathcal{G}}_{\Omega} \left(\begin{smallmatrix}
\left[L_{I,1}(-\Omega,\Omega_2)\right]_{cl,cl}  \\ \left[ L_{I,1}(-\Omega,\Omega_2)\right]_{q,cl}
\end{smallmatrix}\right)
\end{eqnarray}
Finally, we need to perform   shifts of frequencies  in the first term  to write this result in a simple form:
\begin{eqnarray}
\label{S112}
&\left\{\frac{i}{2}\langle [S^{(1)}]^2  \rangle\right\}^{(2)}_f= - 2\left(\frac{g}{ \varepsilon_0}\right)^2 
\int \frac{d\Omega_1d\Omega d\Omega_2}{(2\pi)^3}  \tilde{\Phi}^{cl}_{\Omega_2+\frac{\Omega}{2}} \tilde{\Phi}^{cl}_{-\Omega_2+\frac{\Omega}{2}}\hat{f}_{-\Omega_1-\frac{\Omega}{2}}^T  \hat{\Pi}_{I}(\Omega,\Omega_1,\Omega_2)\hat{f}_{\Omega_1-\frac{\Omega}{2}} \nonumber \\
& \hat{\Pi}_{I}(\Omega,\Omega_1,\Omega_2)=\hat{\Pi}_{I,2}(\Omega,\Omega_1,\Omega_2)+\frac{1}{2}\hat{\Pi}_{I,1}\left(\Omega_1-\Omega_2,-\frac{\Omega_1+\Omega_2+\Omega}{2},\frac{\Omega_1+\Omega_2-\Omega}{2}\right)
\end{eqnarray}

\subsection{Susceptibility: major contribution}
\label{App_4_3}
The major contribution to susceptibility is obtained when we set $\tilde{\Phi}_{\Omega}=2\pi \delta(\Omega)$ in the above
action (\ref{S112}); pumping then enters via modification of the soft-mode action only. The result for the action then reads:
\begin{eqnarray}
\left\{\frac{i}{2}\langle [S^{(1)}]^2  \rangle\right\}^{(2)}_f= - 2\left(\frac{g}{ \varepsilon_0}\right)^2 
\int \frac{d\Omega}{(2\pi)}  \hat{f}_{-\Omega}^T  \hat{\Pi}_{I}(0,\Omega,0)\hat{f}_{\Omega}\nonumber \\ \hat{\Pi}_{I}(0,\Omega,0)=\frac{\Omega^2}{2}\left(\begin{smallmatrix}
0 && \mathcal{G}^{A}_{\Omega} \psi^2(\Omega) \\\mathcal{G}^{R}_{\Omega} \psi^2(-\Omega) && -i\pi \left[\mathcal{G}^{A}_{\Omega} \psi(\Omega)+\mathcal{G}^{R}_{\Omega} \psi(-\Omega)\right] + \mathcal{G}^{K}_{\Omega} \psi(-\Omega)\psi(\Omega)
\end{smallmatrix}\right)
\end{eqnarray}
and susceptibility has now the following  from:
\begin{eqnarray}
\label{chi-box}
\boxed{\chi(\Omega)=  -\frac{2g}{ \varepsilon_0}\psi(-\Omega) \left[1+\frac{g}{ 2\varepsilon_0}
	\Omega^2 \mathcal{G}^{R}_{\Omega} \psi(-\Omega)\right]}
\end{eqnarray}
Here we  have omitted the constant (frequency-independent) term related  with the UV logarithmic divergence at the scale  $\frac{J}{T}$.  We emphasize that Green function in Eq.(\ref{chi-box}) cointains, in general, the effect of pumping.
In the explicit form  the susceptibility is given by
\begin{eqnarray}
\chi(\Omega)=  -\frac{2g}{ \varepsilon_0}\psi(-\Omega)  \frac{\left(\Omega^2+1\right)}{ \left(\Omega^2+1\right)-\frac{g}{2\varepsilon^2_0}\psi(-\Omega)}\nonumber \\ \Im \chi^R(\Omega)=-\frac{2g}{ \varepsilon_0}\Im\psi(-\Omega)  \frac{\left(\Omega^2+1\right)^2}{ \left(\Omega^2+1-\frac{g}{2\varepsilon^2_0}\Re\psi(-\Omega)\right)^2+\left[\frac{g}{2\varepsilon^2_0}\Im\psi(-\Omega)\right]^2}\nonumber \\
\end{eqnarray}

\subsection{Susceptibility: additional term}
\label{App_4_4}
In the presence of the pumping  we can write:  $\tilde{\Phi}_{\Omega}=2\pi\left(\delta(\Omega)+\frac{A}{2}\left(\delta(\Omega-\Omega_P)+\Omega_P\rightarrow-\Omega_P\right)\right)$. Taking into account the terms $\sim A$ in $\tilde{\Phi}_{\Omega}$
results in the additional term of the order $A^2$ in the action 
\begin{eqnarray}
\delta_A\left\{\frac{i}{2}\langle [S^{(1)}]^2  \rangle\right\}^{(2)}_f= - 2\left(\frac{Ag}{ 2\varepsilon_0}\right)^2 
\int \frac{d\Omega }{(2\pi)}   \hat{f}_{-\Omega}^T  \left[\hat{\Pi}_{I}(0,\Omega,\Omega_P)+\Omega_P\rightarrow-\Omega_P\right]\hat{f}_{\Omega} \approx \nonumber \\
i  \frac{\ln(\Omega_R) A^2}{ \Omega_W }  \frac{g}{ \varepsilon_0}
\int \frac{d\Omega }{(2\pi)}   \hat{f}_{-\Omega}^T  \left(\begin{smallmatrix}
0 && -\Omega \\ \Omega && \Omega_P
\end{smallmatrix}\right)\hat{f}_{\Omega}
\end{eqnarray}
which leads  to the following contribution to the susceptibility: 
$\delta_A \chi^R(\Omega)=i  \frac{\ln(\Omega_R) A^2}{ \Omega_W }  \frac{g}{ \varepsilon_0}\Omega$.  It is smaller
than the major contribution as $\Omega/\Omega_W \ll 1$.

\section{Heating}
\label{App_5}
Pumping leads to  absorption of energy and thus to heating of the system. For the SYK model the change of temperature leads to the non-zero average of $\partial_x \langle u^{cl}(x)\frac{1}{\sqrt{2}}\rangle=\frac{T_{new}}{T_{old}}$ so for sufficiently-long heating we will break our assumption $\partial_x u\ll 1$. On the other hand, as it was shown in the main text, dissipation rate for "dry friction" regime does not depend on the temperature; thus our result is  not sensitive to heating,
as long as we neglect the terms in the action  beyond quadratic approximation over soft mode $u(x)$.

To keep the increase of absorbed energy  and heating $\delta T/T$ small, we  consider pumping with small amplitude $A$
and finite duration $t_{\mathrm{pump}}$, which is non-monochromatic by definition.
In this Section we will show how to relate spectrum of our pulse and the change of $u^{cl}$.
We will consider the general case of the pumping in the form $\tilde{\Phi}^{cl}_{\Omega}=\left(2\pi\delta(\Omega)+\Phi_{\Omega}\right)$ and $\Phi^{q}=0$. 
Our aim to write an expression for  $\langle u^{cl}_{\Omega}\rangle$, but first we need to find the propagator of fluctuations of
the soft mode. In  presence of pumping it is no longer  a function of a single frequency, since the problem is non-stationary.

We assume that pumping is weak and  use perturbation theory up to the second order terms in the pumping amplitude.
In the absence of the pumping we have: $\mathcal{G}^{(0)}(\Omega,\Omega^\prime)=2\pi \delta(\Omega+\Omega^\prime) \mathcal{G}_{\Omega}$. where $ \mathcal{G}_{\Omega}$  is determined in the subsection III.A.
The linear correction to the Green function can be obtained from the expression for $S^{(2)}_{\tilde{\Phi}}$ and has the form:
\begin{eqnarray}
\mathcal{G}^{(1)}(\Omega,\Omega^\prime)= \frac{ig}{2 \varepsilon_0} \Phi_{\Omega+\Omega^\prime}\mathcal{G}_{\Omega} \left[\hat{L}_{II}(-\Omega^\prime,-\Omega,\frac{\Omega+\Omega^\prime}{2})+\hat{L}_{II}(-\Omega^\prime,-\Omega,-\frac{\Omega+\Omega^\prime}{2})\right]\mathcal{G}_{-\Omega^\prime}
\end{eqnarray}
Using  Eq.(\ref{S11}) from appendix \ref{App_2_3} for $S^{(1)}$ we find for the average values:
\begin{eqnarray}
\label{average1}
\langle \left(\begin{smallmatrix}
u^{(cl)}_{\Omega} \\  u^{(q)}_{\Omega}
\end{smallmatrix}\right) \rangle = -\frac{2g}{ \varepsilon_0}\int \frac{d\Omega_1d\Omega^\prime}{(2\pi)^2}  \frac{\tilde{\Phi}^{cl}_{-\Omega_1-\frac{\Omega^\prime}{2}} \tilde{\Phi}^{cl}_{\Omega_1-\frac{\Omega^\prime}{2}}}{\sqrt{2}} \mathcal{G}(\Omega,\Omega^\prime) \left(\begin{smallmatrix}
0 \\  1
\end{smallmatrix}\right) L_{I,1}(\Omega^\prime,\Omega_1)_{q,cl}
\end{eqnarray}
We are interesting in the term proportional to the second power of the pumping amplitude $A$.   Since  $S^{(1)} \propto A$,
it is sufficient to use in Eq.(\ref{average1}) the first-order correction $\mathcal{G}^{(1)}$ to the Green function.
Our aim is to calculate 
$$\lim\limits_{\Omega\rightarrow 0 } -\Omega^2\frac{1}{\sqrt{2}}\langle u_{\Omega}\rangle=\partial_x \frac{1}{\sqrt{2}}\langle u(x)\rangle|_{-\infty}^{\infty}=\frac{\delta T}{ T}$$
Using Eq.(\ref{average1}) we find:
\begin{eqnarray}
\label{delta_T_1}
\frac{\delta T}{ T}=-\lim\limits_{\Omega\rightarrow 0 }\langle \frac{\Omega^2}{\sqrt{2}}u^{(cl)}_{\Omega} \rangle=\left[\lim\limits_{\Omega\rightarrow 0 } \Omega^2 G(\Omega) \right]\int \frac{d\Omega^\prime}{2\pi} |\Phi_{\Omega^\prime}|^2\Im\chi(\Omega^\prime) \frac{\Omega^\prime}{2}
\end{eqnarray}
In terms of physical frequency $\omega$ the result reads
\begin{eqnarray}
\label{delta_T_2}
\delta T=\frac{1}{2\pi}\frac{1}{\varepsilon_0+\frac{g}{\varepsilon_0}} \int \frac{d\omega^\prime}{2\pi} |\Phi_{\omega^\prime}|^2
\Im \chi(\omega^\prime) \frac{\omega^\prime}{2}=\frac{Q}{C}
\end{eqnarray}
where
\begin{equation}
Q=\int \frac{d\omega^\prime}{2\pi} |\Phi_{\omega^\prime}|^2\Im\chi(\omega^\prime) \frac{\omega^\prime}{2} 
\label{Q4}
\end{equation}
is the full energy absorption and
\begin{equation}
C= 2\pi\left(\varepsilon_0+\frac{g}{\varepsilon_0}\right)
\label{C2}
\end{equation}
is the heat capacity, compare with  SM, the end of appendix \ref{App_2_2} and Eq.(\ref{C01}).  For  a pumping pulse with frequency
$\omega_R$, amplitude $A$ and duration $t_{\mathrm{pump}}$, we have (with $\delta = 1/t_{\mathrm{pump}}$):
\begin{equation}
\Phi_{\omega} = \frac{A}{2} \left[ \frac{\delta}{\delta^2 + (\omega-\omega_R)^2} + \frac{\delta}{\delta^2 + (\omega +\omega_R)^2} \right]
\label{Phi-omega}
\end{equation}
therefore
\begin{equation}
Q = \frac{A^2}{16}\omega_R \,t_{\mathrm{pump}}\Im\chi(\omega_R)
\label{Ql}
\end{equation}
The results (\ref{Ql},\ref{C2}) were used to derive inequality (24) of the main text.

\section{Modification of the saddle-point solution}
\label{App_6}
In the major part of this paper we have studied the properties of the soft reparametrization  mode  and its impact upon the susceptibility, in presence of quadratic terms $\propto \Gamma_{ij}$.  Below we consider different effect of these terms,
that is,  modification of the saddle-point solution for the fermionic Green function; in other terms, here we account
for the "hard modes" effect.

To analyze hard modes we use the following action: 
\begin{eqnarray}
\label{Action-Int}
S=S_{\varphi}-\frac{i}{2}N\sum_{s,s^{\prime}}\int dxdx^{\prime}\left[\mathcal{M}ss^{\prime}G_{ss^{\prime}}^{2}(x,x^{\prime})+G_{s,s^{\prime}}(x,x^{\prime})\Sigma_{s,s^{\prime}}(x,x^{\prime})\right]-i\frac{N}{2}Tr\ln\left(\hat{1}+\hat{G}^{\varphi}\circ\hat{\Sigma}\right)
\end{eqnarray}
where $\mathcal{M} \equiv \frac{\Gamma^{2}}{2J^{2}}\left(\frac{2\pi}{\beta J}\right)^{-1}$. 
Eq.(\ref{Action-Int}) contains soft-mode dependent Green function 
\begin{eqnarray}
G_{s_1,s_2}^{\varphi}(x_1,x_2)=G_{s_1,s_2}^{0}(\varphi_{s_1}(x_1),\varphi_{s_2}(x_2)) \left[\varphi^\prime_{s_1}(x_1)\varphi^\prime_{s_2}(x_2)\right]^\Delta \nonumber \\
\end{eqnarray}
where $\hat{G}^{0}$ is the  conformal solution for the $SYK$ model,
which has the following form in the $(cl,q)$ basis, in the Fourier domain:
\begin{eqnarray}
\hat{G}=\left(\begin{array}{cc}
\mathcal{G}^{R}(\Omega) & \mathcal{G}^{K}(\Omega)\\
0 & \mathcal{G}^{A}(\Omega)
\end{array}\right)=-2ib^{\Delta}\cos\left(\pi\Delta\right)\left(\begin{array}{cc}
K_{\Delta}(\Omega) & \tanh(\pi\Omega)\left\{ K_{\Delta}(\Omega)+K_{\Delta}(-\Omega)\right\} \\
0 & -K_{\Delta}(-\Omega)
\end{array}\right)
\end{eqnarray}
In the presence of the source field the action has the form:
\begin{eqnarray}
S=S_{\varphi}-\frac{i}{2}N\sum_{s,s^{\prime}}\int dxdx^{\prime}\left[\mathcal{M}\left(1+\Phi^{s}(x)\right)\left(1+\Phi^{s^{\prime}}(x^{\prime})\right)ss^{\prime}G_{ss^{\prime}}^{2}(x,x^{\prime})+G_{s,s^{\prime}}(x,x^{\prime})\Sigma_{s,s^{\prime}}(x,x^{\prime})\right]-\nonumber \\-i\frac{N}{2}Tr\ln\left(\hat{1}+\hat{G}^{\varphi}\circ\hat{\Sigma}\right) \nonumber \\
\end{eqnarray}
It is useful to rewrite it using matrix notations:
\begin{eqnarray}
&S=S_{\varphi}+\frac{i}{2}N\mathcal{M}Tr\left[\left\{ \hat{1}+\hat{\mathcal{F}}\right\} \circ G\circ\left\{ \hat{1}+\hat{\mathcal{F}}\right\} \circ G\right]+\frac{i}{2}NTr\left[\hat{G}\circ\hat{\Sigma}\right]-i\frac{N}{2}Tr\ln\left(\hat{1}+\hat{G}^{\varphi}\circ\hat{\Sigma}\right) \nonumber \\
&\hat{\mathcal{F}}(x,x^{\prime})=\left(\begin{array}{cc}
\Phi^{cl}(x) & \Phi^{q}(x)\\
\Phi^{q}(x) & \Phi^{cl}(x)
\end{array}\right)\delta(x-x^{\prime})
\end{eqnarray}
In the limit $\mathcal{M}\ll1$ fluctuations near the saddle point of the field $G$ and $\Sigma$ are small. Let us consider
the effect of these fluctuations assuming that $G=G^{\varphi}+\delta G$ and $\Sigma=\delta \Sigma$.
In this case:
\begin{eqnarray}
S=S_{\varphi}+\frac{i}{2}N\mathcal{M}Tr\left[\left\{ \hat{1}+\hat{\mathcal{F}}\right\} \circ\hat{G}^{\varphi}\circ\left\{ \hat{1}+\hat{\mathcal{F}}\right\} \circ\hat{G}^{\varphi}\right]+\nonumber\\
i N\mathcal{M}Tr\left[\left\{ \hat{1}+\hat{\mathcal{F}}\right\} \circ\delta\hat{G}\circ\left\{ \hat{1}+\hat{\mathcal{F}}\right\} \circ\left(\hat{G}^{\varphi}\right)\right] +\frac{i}{2}N\mathcal{M}Tr\left[\left\{ \hat{1}+\hat{\mathcal{F}}\right\} \circ\delta\hat{G}\circ\left\{ \hat{1}+\hat{\mathcal{F}}\right\} \circ\delta\hat{G}\right]\nonumber\\
+\frac{i}{2}NTr\left[\left(\hat{G}^{\varphi}+\delta\hat{G}\right)\circ\hat{\Sigma}\right]-i\frac{N}{2}Tr\left(\hat{G}^{\varphi}\circ\delta\hat{\Sigma}-\frac{1}{2}\hat{G}^{\varphi}\circ\delta\hat{\Sigma}\circ\hat{G}^{\varphi}\circ\delta\hat{\Sigma}\right)
\end{eqnarray}
The first line  is the action of our original model (soft-mode only)  and we had calculated 
the corresponding susceptibility already.  The second term in the second line is unimportant since 1) this correction to the action is small as $\mathcal{M}\ll1$, and 2) it contains two $\delta G$ fields and thus the correction to susceptibility will contain
extra small factor $1/ N \ll 1$.  We also note that in the leading order there is no mixing  between soft mode $\varphi$ and fields $G$ and $\Sigma$. As a result the correction to  susceptibility  is determined by the action:
\begin{eqnarray}
\delta S\left[\frac{i}{2}N\right]^{-1}=2\mathcal{M}Tr\left[\delta\hat{G}\circ\left\{ \hat{1}+\hat{\mathcal{F}}\right\} \circ\left(\hat{G}^{0}\right)\circ\left\{ \hat{1}+\hat{\mathcal{F}}\right\} \right]+Tr\left[\delta\hat{G}\circ\hat{\Sigma}\right]+\frac{1}{2}Tr\left(\hat{G}^{\varphi}\circ\hat{\Sigma}\circ\hat{G}^{\varphi}\circ\hat{\Sigma}\right) \nonumber\\
\end{eqnarray}
After integration over $\delta G$ and $\delta \Sigma$ we have the following action :
\begin{eqnarray}
\label{Shard}
S_{\mathcal{F}}\left[\frac{i}{2}N\right]^{-1}=2\left(2\mathcal{M}\right)^{2}Tr\left[\hat{\mathcal{F}}\circ\hat{G}^{0}\circ\hat{\mathcal{F}}\circ\hat{G}^{0}\circ\hat{G}^{0}\circ\hat{G}^{0}\right]+\nonumber\\\left(2\mathcal{M}\right)^{2}Tr\left[\hat{\mathcal{F}}\circ\hat{G}^{0}\circ\hat{G}^{0}\circ\hat{\mathcal{F}}\circ\hat{G}^{0}\circ\hat{G}^{0}\right]
\end{eqnarray}
The action (\ref{Shard}) can be used to calculate correction to the susceptibility; the result is
\begin{eqnarray}
\label{delta-chi-hard}
\delta \chi(\Omega)=-i4N\mathcal{M}^{2}\int\frac{d\Omega^{\prime}}{2\pi}F_{\Omega^{\prime}}(\left\{ G_{\Omega^{\prime}}^{R}-G_{\Omega^{\prime}}^{A}\right\} \left\{ \left[G_{\Omega^{\prime}+\Omega}^{R}\right]^{3}+\left[G_{\Omega^{\prime}-\Omega}^{A}\right]^{3}\right\} \nonumber\\+\left\{ \left[G_{\Omega^{\prime}}^{R}\right]^{3}-\left[G_{\Omega^{\prime}}^{A}\right]^{3}\right\} \left\{ G_{\Omega^{\prime}+\Omega}^{R}+G_{\Omega^{\prime}-\Omega}^{A}\right\} +\left\{ \left[G_{\Omega^{\prime}}^{R}\right]^{2}-\left[G_{\Omega^{\prime}}^{A}\right]^{2}\right\} \left\{ \left[G_{\Omega^{\prime}+\Omega}^{R}\right]^{2}+\left[G_{\Omega^{\prime}-\Omega}^{A}\right]^{2}\right\} )
\end{eqnarray}
Taking the imaginary part of Eq.(\ref{delta-chi-hard}) in the low-$\Omega$ limit, we obtain Eq.(23) of the main text.

\end{appendix}



\bibliography{Paper_to_SciPost_BiBTeX_File}

\end{document}